\documentclass[usenatbib,useAMS,usegraphicx]{mn2e}

\usepackage{xspace}
\usepackage[fleqn]{amsmath}
\usepackage{times}

\providecommand{\eg    }{e.g.\xspace}%
\providecommand{\etal  }{et al.\@\xspace}%
\providecommand{\ie    }{i.e.\xspace}
\providecommand{\xray  }{X-ray\xspace}%
\providecommand{\xrays }{X-rays\xspace}%
\providecommand{\gray  }{$\gamma$-ray\xspace}%
\providecommand{\grays }{$\gamma$-rays\xspace}%
\providecommand{\fermi }{\textit{Fermi}\xspace}%
\providecommand{\chandra}{\textit{Chandra}\xspace}%
\providecommand{\swift}{\textit{Swift}\xspace}%
\providecommand{\rxte}{\textit{RossiXTE}\xspace}%

\newcommand\aj{\textrm{AJ}}%
\newcommand\araa{\textrm{ARA\&A}}%
\newcommand\apj{\textrm{ApJ}}%
\newcommand\apjl{\textrm{ApJ}}%
\newcommand\apss{\textrm{Ap\&SS}}%
\newcommand\aap{\textrm{A\&A}}%
\newcommand\mnras{\textrm{MNRAS}}%
\newcommand\pasp{\textrm{PASP}}%
\title[Variability of external Compton and SSC models]%
{Time-dependent simulations of emission from FSRQ PKS~1510$-$089:
multiwavelength variability of external Compton and SSC models}

\author[X. Chen et al.]{%
Xuhui~Chen,$^1$\thanks{gfossati@rice.edu, xuhui.chen@alumni.rice.edu}
Giovanni~Fossati,$^1$\footnotemark[1] 
Markus B\"ottcher$^2$, and Edison Liang$^1$ \\
$^1$ Department of Physics and Astronomy, Rice University, Houston, TX 77005 \\
$^2$ Astrophysical Institute, Department of Physics and Astronomy, Ohio University, Athens, OH 45701
}

\begin{document}

\date{Accepted 2012 May 9.  Received 2012 May 8; in original form 2012 March 31}

\pagerange{\pageref{firstpage}--\pageref{lastpage}} \pubyear{2012}

\maketitle

\label{firstpage}

\begin{abstract}
We present results of modeling the broadband SED and multiwavelength
variability of the bright FSRQ PKS~1510$-$089 with our time-dependent multizone
Monte Carlo/Fokker-Planck code \citep{chen_etal:2011:multizone_code_mrk421}. 
As the primary source of seed photons for inverse Compton scattering, we
consider radiation from the broad line region (BLR), from the hot dust of the
molecular torus, and the local synchrotron radiation (synchrotron-self-Compton,
SSC). 
We evaluate the viability of different Compton models by comparing simulated
multiwavelength light curves and SEDs with one of the best observed flares 
by PKS~1510$-$089, in March 2009.
The time-dependence of our code and its correct handling of light travel time
effects allow us to fully take into account the effect of the finite size of
the active region, and in turn to fully exploit the information carried by
time resolved observed SEDs that are becoming increasingly available since the
launch of \fermi.
We confirm that the spectrum adopted for the external radiation field has an
important impact on the modeling of the SED, in particular for the lower energy
end of the Compton component which is observed in the \xray band, which in turn
is one of the most critical bands to assess the differences between EC and SSC
emission.
In the context of the scenario presented in this paper, where the flaring is
caused by the increase of the number of relativistic electrons ascribed to the
effect of the interaction of a portion of the jet (blob) with a shock, we can
not firmly discriminate the three main scenarios for \gray emission.
However, results show clearly the differences produced by a more realistic
treatment of the emitting source in the shape of SEDs and their time
variability over relevant, observable time-scales, and demonstrate the
crucial importance of time-dependent multi-zone models to advance our
understanding of the physics of these sources, by taking full advantage
of the wealth of information offered by the high quality data of current
multiwavelength campaigns.
\end{abstract}


%%%%%%%%%%%%%%%%%%%%%%%%%%%%%%%%%%%%%%%%%%%%%%%%%%%%%%%%%%%%%%%%%%%%%%%%%%%%%%%%
\section{Introduction}
\label{sec:introduction}

The spectral energy distributions (SED) of blazars usually show two major 
non-thermal components. The low energy one, peaking in the IR -- optical -- 
\xray range is identified as synchrotron radiation. 
The origin of the high energy one, peaking in the \gray energy range (MeV to
TeV) is less clear.
Proposed ideas include leptonic models, based on inverse Compton (IC)
scattering by the same electrons emitting the synchrotron radiation, and
hadronic models, in which protons play a critical role in producing the high
energy emission \citep{mannheim:1998:science_cosmic_rays,
rachen_correlated_x_tev,
sikora_madejski:2001:review,
arbeiter_etal:2005:ssc_and_pions,
levinson:2006:review,
boettcher:2007:emission_processes_review,
dermer_lott:2011:leptonic_models_for_blazars}.
For the leptonic IC-based models, several sources of the target photons are
possible and the debate about which one is dominant and in which type of object
has recently been reignited.
If the seed photons are provided by the synchrotron radiation emitted at lower
energy by the same IC-scattering electrons, it is referred to as synchrotron
self-Compton (SSC).
Scenarios in which the dominant contribution to the seed soft photon field for
IC is provided by radiation emitted elsewhere are referred to as external
Compton (EC) models. 
External sources of seed photons may include the photons from accretion disc,
hot \xray-emitting corona, broad emission line region, infrared torus,
host galaxy bulge, and cosmic background radiation \citep{gg_tavecchio:2009:canonical_blazars}.

There are two major classes of blazars: BL Lac objects, which have featureless
optical spectra, and Flat Spectrum Radio Quasars (FSRQ), which exhibit broad
quasar-like emission lines.
The presence of these latter in FSRQs suggests that their jets are in an
environment with a stronger external radiation field. 
Furthermore, depending on the relative location of these sources external to
the jet and of active jet region (blob), the emission from the external
sources would be relativistically beamed and enhanced in the frame of the
blob, possibly making them dominant over the locally produced synchrotron
emission.  
Therefore, the EC model is frequently invoked to explain the emission of 
FSRQs \citep{dermer_etal:1992,
sikora_begelman_rees:1994:external_compton,
sikora_etal:2009:constraining_emission_models}.

A major defining feature of blazars is their rapid and large variability,
observed over the entire range of their continuum emission (radio to TeV
\grays).  
Simultaneous multiwavelength observations and the correlation analysis of the
observed multiwavelength variability can provide insights to the physics of
particle processes and radiation mechanisms in the jet.
The detailed observation and modeling of such variability has been performed
extensively for High energy peaked BL Lacs (HBL) such as Mrk~421 
\citep{fossati_etal:2008:xray_tev,chen_etal:2011:multizone_code_mrk421} 
and PKS~2155$-$304 \citep{aharonian_etal:2007:pks2155_exceptional_flare,
katarzynski_etal:2008:pks2155}.
In the case of HBLs, such studies are facilitated by their synchrotron SED peak
falling in the \xray range which can be observed with multiple \xray
satellites, while their high energy SED peak in TeV \gray is covered by ground
based Atmospheric Cherenkov Telescopes (for a review, \citealp{hinton_hofmann:2009:ARAA};
examples for Mrk~421, \citealp{fossati_etal:2008:xray_tev,abramowski_etal:2012:mrk421_HESS_2006}).
  
The launch of \fermi has re-opened the GeV \gray sky with unprecedented
sensitivity and daily coverage. This energy band covers a highly variable part
of the SED right above the peak of the high energy component of the SED of
several bright FSRQs, such as PKS~1510$-$089
\citep{abdo_etal:2010:pks1510_multiwavelength_flares_2008_2009} 
and 3C454.3 \citep{abdo_etal:2009:3C454}. 
Simultaneous coverage in other wavelength such as optical and \xrays provided
us a chance to obtain multi-epoch SEDs and cross-band correlations. 
A deeper understanding of these time series data sets requires time-dependent
modeling with all light travel time effects (LTTEs) taken into account.

The importance of the LTTE in the study of blazars has long been realized
\citep{chiaberge_ghisellini:1999:timedep}. 
The observed change of flux level on time-scale of hours in some blazars
indicates that these effects must have a large impact on the variability of
blazars. 
There have been some efforts to include these effects in the modeling of blazars 
\citep{kataoka_etal:2000:crossingtimes_model,
sokolov_marscher_mchardy:2004:SSC,
sokolov_marscher:2005:EC,
katarzynski_etal:2008:pks2155,
graff_etal:2008:pipe}, but none of them have taken into account all of these
effects in a generic 2D geometry.

We have developed a time-dependent multizone code using the Monte
Carlo method for radiation transport and the Fokker-Planck equation 
for electron evolution, which we first applied to a study of the correlated
\xray/\gray variability of the HBL Mrk~421 using a pure SSC model
\citep{chen_etal:2011:multizone_code_mrk421}. 
In this paper we extend the model to include external sources of IC seed
photons and apply it to study the multiwavelength variability of an
archetypical powerful FSRQ, PKS~1510$-$089.

%%%%%%%%%%%%%%%%%%%%%%%%%%%%%%%%%%%%%%%%%%%%%%%%%%%%%%%%%%%%%%%%%%%%%%%%%%%%%%%%
\section{PKS~1510$-$089}
\label{sec:pks1510}

PKS~1510$-$089 is a FSRQ at a redshift of $z = 0.361$ \citep{thompson_etal:1990}.
It is one of the brightest and most variable sources detected by \fermi/LAT.
A feature that can be interpreted as disk emission (big blue bump, BBB) is
clearly visible in its optical/UV spectrum.  
VLBI observations of its jet show superluminal motion with apparent speed up to
$45 c$ \citep[$\langle\Gamma\rangle=36$,][]{jorstad_etal:2005:VLBA}.

Since the advent of \fermi, the long-term multiwavelength monitoring
effort, complemented by more intense campaigns motivated by flaring phases, has 
lead to the observation of several large correlated flares for PKS~1510$-$089 
\citep{abdo_etal:2010:suzaku_obs_of_FSRQs,
abdo_etal:2010:pks1510_multiwavelength_flares_2008_2009,
dammando_etal:2011:pks1510_with_agile_march_2009,
marscher_etal:2010:pks1510,
kataoka_etal:2008:pks1510,
pucella_etal:2008:pks1510}.

As reference data for our simulation we choose the observations of a high state
observed in 2008--2009 and presented by 
\citet{abdo_etal:2010:pks1510_multiwavelength_flares_2008_2009}, 
\citet{dammando_etal:2011:pks1510_with_agile_march_2009} and 
\citet{marscher_etal:2010:pks1510}.
One particular flare at the end of March 2009 (peaking around March 25th, MJD
54917) is chosen as the benchmark for this study.

We aim to reproduce several observational features by matching both
the simulated light curves and SED with the observed ones. 
These features include:
\begin{itemize}

\item Clear presence of BBB emission with the general two non-thermal
continuum components SED.  The BBB is evident in both the high and the
low states of the jet non-thermal continuum emission.

\item The time-scale of the flares, which were typically about 4 days (300~ks)
at all wavelengths.

\item Infrared and \gray light curves show the stronger variations among the
observed bands, with similar amplitude, up to a factor of 10.
In the March 2009 flare, the infrared (R band) and \gray (\fermi/LAT)
fluxes were strongly correlated, with no significant lags.

\item The variations in the \swift/UVOT bands were less prominent than 
those in the infrared -- optical bands.
The optical/UV spectral shape became softer when the source brightness increased,
consistent with the combination of a variable softer broad band continuum
(synchrotron) and a non variable component peaked in the UV band (BBB).

\item The variability in the \xray band luminosity was modest, within a factor
of 2 over a period of 4 months encompassing the flare that we selected for this
study (\swift/XRT in \citealp{abdo_etal:2010:pks1510_multiwavelength_flares_2008_2009},
\rxte/PCA in \citealp{marscher_etal:2010:pks1510}).
The spectrum was always very hard, with \textit{energy} index $\alpha_{X}<0.6$
($F_\nu \propto \nu^{-\alpha}$) with very little variability.  
During the period around the benchmark flare it remained $\alpha_{X}\simeq0.5$ 
\citep{abdo_etal:2010:pks1510_multiwavelength_flares_2008_2009}.

\item The $>\!0.2$~GeV spectral shape as measured by \fermi/LAT did not vary
significantly through large luminosity changes, remaining around $\alpha_\gamma
\simeq 1.5$ (for a power law fit)
\citep{abdo_etal:2010:pks1510_multiwavelength_flares_2008_2009}. 

\end{itemize}
These characteristics only coarsely summarize the true richness of information
provided by the full multiwavelength and multi-epoch data set which thanks to
our simulations we can try to exploit more deeply.
Nevertheless, because they constitute a quicker and easier way of guiding 
the setup and evaluation of the simulations and we will refer to them when
discussing the comparison of our simulation results with observations in the
following sections.

%---------------------------------------------------------------------
\begin{figure}
\centerline{%
\includegraphics[width=0.99\linewidth]{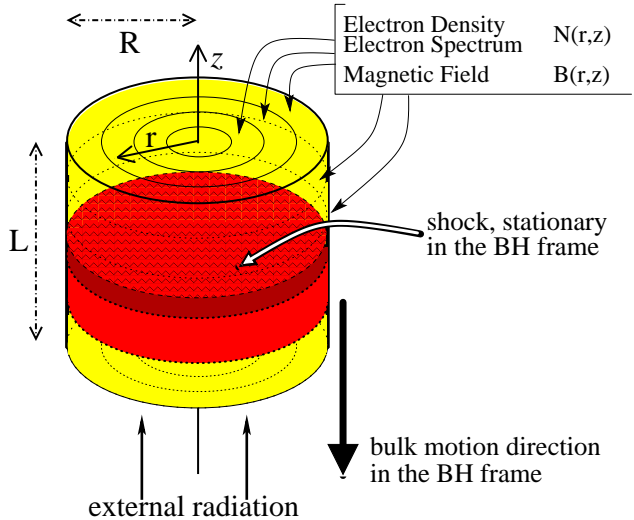}
}
\caption{
The geometry of the model. 
The volume is divided in the $r$ and $z$ directions in zones with their own
electron distribution and magnetic field.  
We also schematically show the setup for the variability of the simulations
with a shock.
The hatched layer represents a stationary shock.
The blob moves downward and crosses the shock front.
For illustration purposes, we plot in lighter color shade the zones that
crossed the shock at earlier times and have had some time to radiate the newly
injected energy.
In this representation the photons from the external radiation fields, beamed
in the blob rest frame, enter the blob from the bottom surface.
\label{fig:geometry}
}
\end{figure}
%---------------------------------------------------------------------

%-----------------------------------------------------------------------
\begin{table*}
\begin{minipage}{\linewidth}
% \tiny
% \scriptsize
\tabcolsep 4pt
\caption{Summary of model parameters}
\label{tab:model_parameters}
\begin{tabular*}{\linewidth}{@{\extracolsep{\fill}}| l | cccccc |}
%-------
\hline
Parameter &   
\texttt{nf/blr} &
\texttt{blr15} & 
\texttt{blr15highgmin} & 
\texttt{blr25} &  
\texttt{torus15} &
\texttt{ssc} \\
\hline                                                                                                                                                  
 Jet bulk Lorentz factor, $\Gamma$                                 &     15.0      &      15.0  &    15.0    &     25.0      &     15.0       &     10.0   \\
 $Z$ (10$^{16}$ cm)                                                &      8.0      &       8.0  &     8.0    &    13.33      &      8.0       &      5.0   \\
 $R$ (10$^{16}$ cm)                                                &      6.0      &       6.0  &     6.0    &     10.0      &      6.0       &     3.75   \\
 Magnetic field, $B$                                               &      0.3      &       0.3  &     0.3    &     0.16      &      0.2       &      0.1   \\
 Particle density (initial), $n_\mathrm{e}$ (10$^{4}$ cm$^{-3}$)   &      2.66     &      2.66  &    2.66    &     0.14      & 7.37 10$^{-2}$ &     0.01   \\ 
\hline                                                                             
 Particle escape time-scale, $t'_\mathrm{esc}$ ($Z/c$)             &      0.1      &       0.1  &     0.1    &      0.1      &    0.015       &     0.03   \\
 Particle (diffuse) acceleration time-scale, $t'_\mathrm{acc}$ ($Z/c$) &     0.55      &      0.55  &    0.55    &     0.55      &     0.09       &     0.19   \\
\hline                                                                             
 Electron pick up rate, Q$_\mathrm{pick}$ (cm$^{-3}$ s$^{-1}$)     &      0.1      &       0.1  &     0.1    & 3.2 10$^{-3}$ &   0.0191       &    0.002   \\
 Pick-up electrons energy, $\gamma_\mathrm{pick}$                  &      5.0      &       5.0  &     5.0    &      5.0      &     50.0       &   1200.0   \\
\hline                                                                             
 Shock injection: $\gamma_\mathrm{min,inj}$                        &      ...      &        30  &      90    &        6      &      300       &     2000   \\
 Shock injection: $\gamma_\mathrm{max,inj}$                        &      ...    & 2 10$^{4}$ & 2 10$^{4}$ & 4 10$^{3}$    & 2 10$^{5}$     &  10$^{5}$  \\
 Shock injection: power-law slope, $p_\mathrm{inj}$                &      ...      &       3.2  &     3.2    &      3.2      &      3.2       &      3.2   \\
 Shock injection: rate: $L'_\mathrm{inj}$ (10$^{44}$ erg s$^{-1}$) &      ...      &       3.5  &     2.0    &      2.8      &      5.0       &      8.0   \\
\hline                                                                           
 $R_\mathrm{BLR}$ or $R_\mathrm{IR}$ (10$^{18}$ cm)                &      0.8     &     0.8    &     0.8    &      0.8      &      7.8       &      ...   \\
 $f_\mathrm{BLR}$ or $f_\mathrm{IR}$                               &      0.013   &     0.013  &     0.013  &      0.0015   &      0.5       &      ...   \\
\hline
\end{tabular*}
\end{minipage}
\end{table*}
%-------------------------------------------------------------------------------

%%%%%%%%%%%%%%%%%%%%%%%%%%%%%%%%%%%%%%%%%%%%%%%%%%%%%%%%%%%%%%%%%%%%%%%%%%%%%%%%
\section{Simulations}
\label{sec:simulations}

\subsection{Basic setup and model parameters}
\label{sec:setup}

Details and technical aspects of our Monte Carlo/Fokker-Planck code are
described in \citet[][\citealt{chen:2012:PhD_thesis}, Chen \etal in preparation]{chen_etal:2011:multizone_code_mrk421}. 
The code uses the Monte Carlo method to track the production, travel, and
Compton scattering of photons, while it solves the isotropic Fokker-Planck
equation to follow the evolution of electrons. 
The major strength and unique feature of this code is that it takes into
account all the LTTEs, internal to the source volume (\eg important for IC
emission) and external, \ie their effect on the observed radiation which we
will receive at different times depending on where it was emitted in the source.

We model a jet active region (blob) as a cylindrical volume crossing a standing
`shock' as illustrated in Fig.\ref{fig:geometry}. 
In the blob rest frame, where all calculations are performed, the shock moves
through the cylindrical region with a speed equal to the bulk velocity of the
blob $v_{bulk} \sim c$.
The cylindrical volume is divided evenly into zones in the radial and vertical
directions ($r$ and $z$ coordinates, $n_r$, $n_z$). 
In all runs presented in this paper, $n_r=9$ and $n_z=30$.  

At this stage the meaning of this shock is simply that of an agent affecting
the properties of the simulation zone where it is at a given time. 
In the simulations presented in this paper it affects the electron distribution,
namely it injects high energy particles into the zones it currently resides in.
It does not affect the magnetic field, which is assumed to be homogeneous
throughout the volume and non varying.
The shock has not thickness (however, our resolution is limited by $dz$) and is
treated as a surface perpendicular to the $z$ direction. 

With respect to the simulations presented in \citep{chen_etal:2011:multizone_code_mrk421},
we added a few features to the code. 
The most relevant ones are the inclusion of a spatially diffuse stochastic
acceleration process, similar to the one discussed by \citet{katarzynski_etal:2006:stochastic},
of a particle escape term and of a particle `pick-up' term 
\citep[see also][]{tramacere_etal:2011:stochastic_acceleration}.
These developments and their motivation are discussed in
\citet{chen_etal:2011:guangzhou}, \citet{chen:2012:PhD_thesis} and 
\citet{chen_etal:2012:hard_xray_lags}.
We treat stochastic acceleration in the whole blob as a diffusive term
in the Fokker-Planck equation, while the putative first order Fermi
acceleration at the shock front is simplified as directly injecting high energy
particles with a power law distribution into the zones where the shock
is present.  

The main parameters of the model are (see also Table~\ref{tab:model_parameters}):
\begin{itemize}
\item $\Gamma$, the bulk Lorentz factor of the jet.

\item $R$ and $Z$, the radius and height of the cylindrical simulation region.

\item $B$, the magnetic field strength, assumed here to be homogeneous and non variable.

\item $n_\mathrm{e}$, the initial electron number density.

\item $t'_\mathrm{acc}$ and $t'_\mathrm{esc}$, the time-scales parameterizing
the stochastic diffuse acceleration and particle escape.

\item $Q_\mathrm{pick}$, the rate at which the blob constantly picks up mildly
relativistic electrons (with a narrow Gaussian distribution centered at
$\gamma_\mathrm{pick}$.

\item $L'_\mathrm{inj}$, the luminosity of the relativistic electrons
injected by the shock.
  
\item $p_\mathrm{inj}$, $\gamma_\mathrm{min,inj}$ and $\gamma_\mathrm{max,inj}$
are the spectral index, minimum and maximum Lorentz factor of the electrons
injected locally by the shock with a power law spectrum.

\item $R_\mathrm{BLR}$, $f_\mathrm{BLR}$ are the size of the broad line region,
assumed to be spherical and the fraction of the luminosity from the accretion
disk that contributes to the radiation energy density within its volume.
$R_\mathrm{IR}$ and $f_\mathrm{IR}$ are the corresponding parameters for the
case of the infrared emitting torus. Collectively we also call them 
$R_\mathrm{ext}$ and $f_\mathrm{ext}$ without distinction between BLR and IR.  
We will discuss them in the following sections.

\end{itemize}

At the beginning of a simulation each zone of the blob is filled with electrons
with density $n_\mathrm{e}$ and a power law spectrum with slope and energy range
given by $p$, $\gamma_\mathrm{min}$ and $\gamma_\mathrm{max}$.
The same parameters are used in every zone.
These parameters are not listed in Table~\ref{tab:model_parameters} because
their relevance is minimal; they are simply seeding each zone with electrons at
beginning of the simulation, and they do not represent the electron
distribution once the simulations starts.  
The actual electrons distribution in each zone will be different from the
seeded one, and from zone to zone, as they evolve separately.
In the steady state it would approximate a power-law like distribution, with
$\gamma_\mathrm{min}$ determined by the energy of the picked-up electrons.
The spectral index is related to $t'_\mathrm{acc}$ and $t'_\mathrm{esc}$,
with faster escape and slower acceleration generally leading to softer
spectrum. 
The $\gamma_\mathrm{max}$ is determined by the competition between acceleration
and cooling, with faster cooling meaning smaller maximum electron energy.

%=======================================================================
\subsection{External Radiation}
\label{sec:ext_rad}

The relativistic jets in Active Galactic Nuclei (AGN) may reside in dense
external radiation environments, with contributions from the BLR 
\citep{tavecchio_gg:2008:blr_and_blazars,poutanen_stern:2010:GeV_breaks} or
the warm dust of the molecular torus hypothesized to exist beyond the accretion
disk \citep{malmrose_etal:2011:hotdust_emission_in_gammaray_bright_blazars}.
For a thorough discussion we refer the reader to \citet{gg_tavecchio:2009:canonical_blazars}.

The dominance of different sources of external radiation is connected to 
the location of the \gray emitting region within the jets. The radiation 
from the BLR can be dominant only when the emission region is located at 
sub-parsec distance from the central engine of the AGN. Beyond that 
distance, the infrared radiation from the dust torus is likely to dominate 
on parsecs scale \citep{gg_tavecchio:2009:canonical_blazars}.
\citet{poutanen_stern:2010:GeV_breaks} argue that the GeV spectral breaks
of FSRQs observed by \fermi/LAT are a sign of \gray absorption inside the BLR.
Meanwhile, \citet{marscher_etal:2010:pks1510} used the correlation between
radio knot appearance and \gray flares to identify the location of the
emission region at several parsecs from the central engine.
We will test the viability of both of these two sources of external photons,
and see if they can produce the SEDs and light curves observed.

We regard the big blue bump clearly visible in the SED of PKS~1510$-$089 when
in its lower brightness states as unbeamed thermal emission from the accretion
disc, and match the data with a luminosity of $4\times10^{45}$ ergs/s and a
temperature of $3\times10^4$~K.  

This disc emission is used to estimate the energy density experienced in the
blob rest frame while its location is within the radius of BLR, $R_\mathrm{BLR}$, 
according to the following transformation equation \citep{gg_madau:1996}:
\begin{equation}
\label{eq:u_blr}
U'_\mathrm{BLR} \sim \frac{17}{12} \frac{f_\mathrm{BLR}\, L_\mathrm{d}\, 
\Gamma^2}{4 \pi R_\mathrm{BLR}^2 c}.
\end{equation}
where $L_\mathrm{d}$ is the disc luminosity.

Similarly, assuming for simplicity that the region where radiation from the
dusty torus yields a significant energy density can be described as a spherical
volume of radius $R_\mathrm{IR}$, the energy density within this region as 
seen in the blob rest frame can be estimated by 
\begin{equation}
\label{eq:u_ir}
U'_\mathrm{IR} \sim \frac{f_\mathrm{IR}\, L_\mathrm{d}\, \Gamma^2}{4 \pi 
R_\mathrm{IR}^2 c}
\end{equation}
\citep{gg_tavecchio:2009:canonical_blazars}. 

For the spectrum of BLR emission, we consider two cases: an approximation as a
plain single temperature blackbody peaked at $1.5 \Gamma \nu_{Ly\alpha}$, or 
a more realistic spectrum obtained by taking the unbeamed BLR spectrum in Fig.~4 of 
\citet{tavecchio_gg:2008:blr_and_blazars} and beam it according to the equation:
\begin{equation}
\label{eq:beaming}
U'(\nu')=\frac{2\pi}{\Gamma\beta c}\nu'^2 \int_{\nu_1}^{\nu_2}
\frac{I(\nu)}{\nu^3} \mathrm{d}\nu,
\end{equation}
here $\nu_1=\nu'/[\Gamma(1+\beta)]$, $\nu_2=\nu'/\Gamma$, $I(\nu)$ is the
unbeamed intensity spectrum, 
$\nu$ and $\nu'$ are the frequency 
in the observer's frame and blob frame respectively.

For the infrared emission from the hot dusty torus, we use a blackbody spectrum
with temperature \citep{gg_tavecchio:2009:canonical_blazars}:
\begin{equation}
\label{eq:T_ir}
T'_{IR} \;=\; 370\; \Gamma\; (1-\beta \cos\alpha)~\text{K} \sim 370\; \Gamma~\text{K},
\end{equation}
where $\alpha$ is the angle
between the jet axis and the line connecting the source and the jet, so
$\alpha \sim \pi/2$ for the torus.

For computational ease, we simplify the model by assuming that all the external
photons are traveling in the upward direction in the frame of the blob,
as illustrated in Fig.~\ref{fig:geometry}. 
All the external photons enter the blob through the lower boundary and 
the external flux is just the energy density times $c$. 
This is a valid approximation for the typical (large) values of the Lorentz
factor appropriate for blazars, which we adopt for ease of computation,
although our code allows to setup the flux from the external illuminating
source with an angular distribution.  

To produce the observed SEDs the disc emission is added as a non-varying
component to the beamed emission from the jet in the post-processing of the
simulation results.

%-----------------------------------------------------------------------
\begin{figure*}
\centerline{
\includegraphics[width=0.32\linewidth]{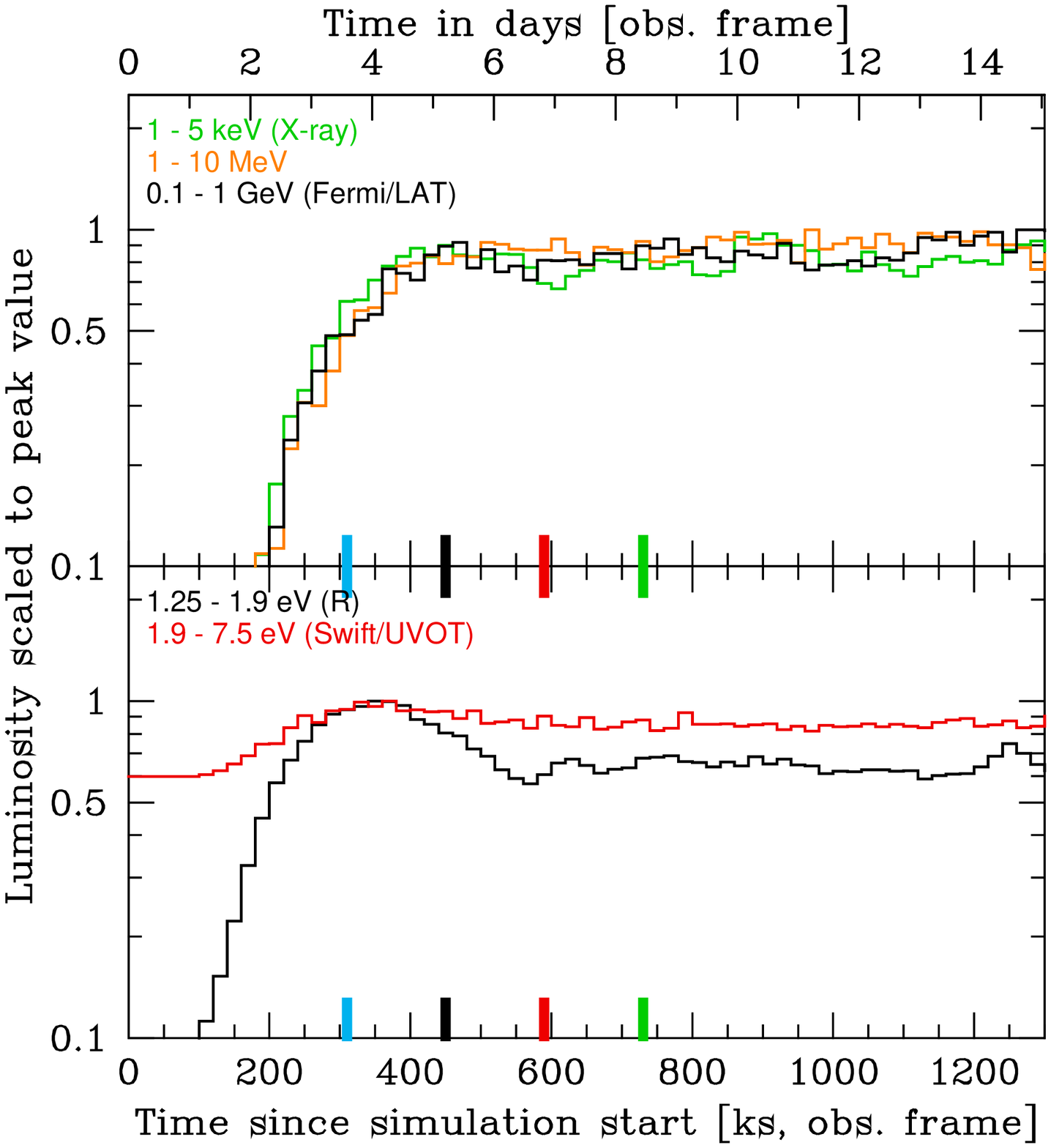}
\includegraphics[width=0.32\linewidth]{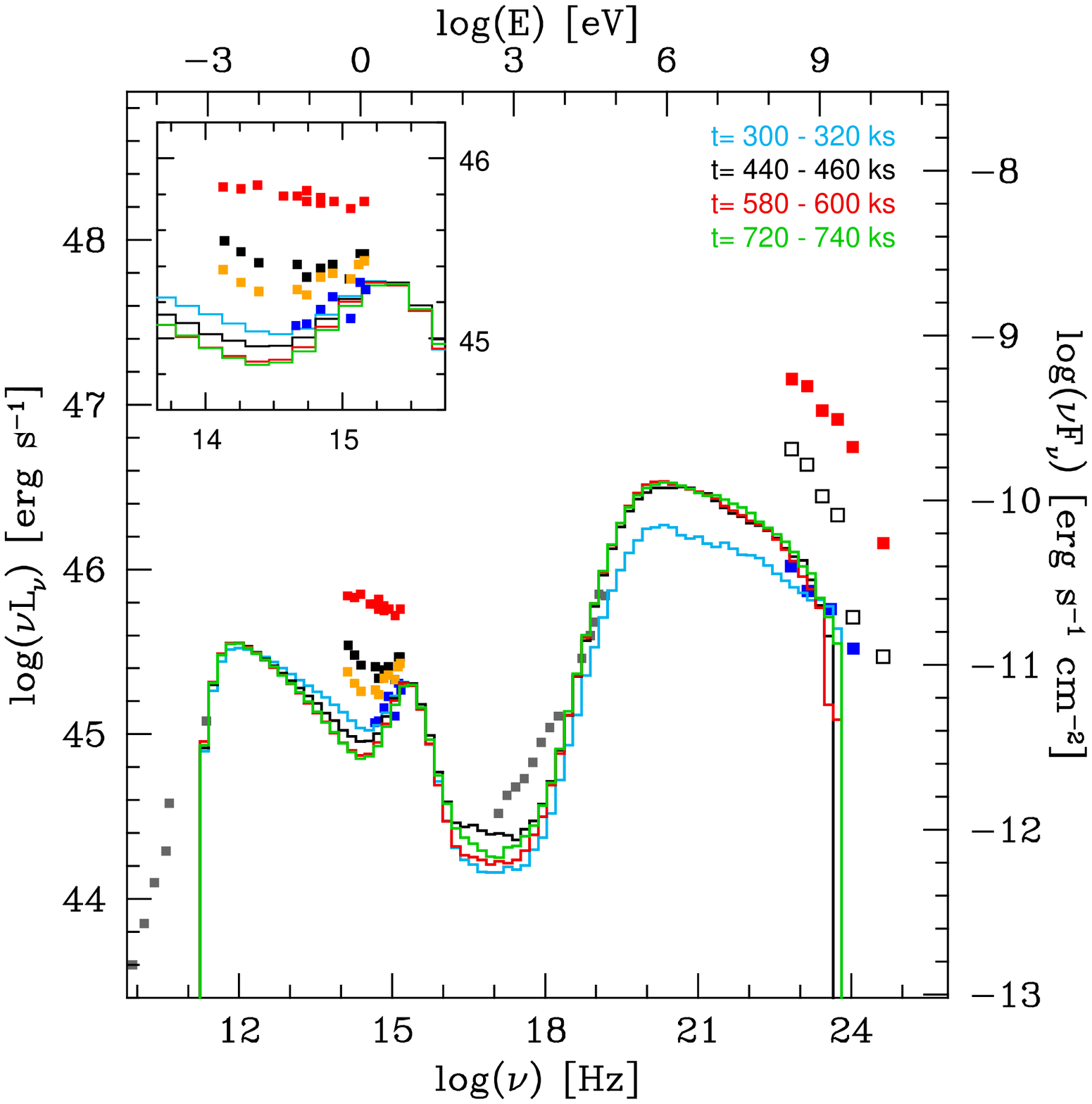}
\hfill
\includegraphics[width=0.32\linewidth]{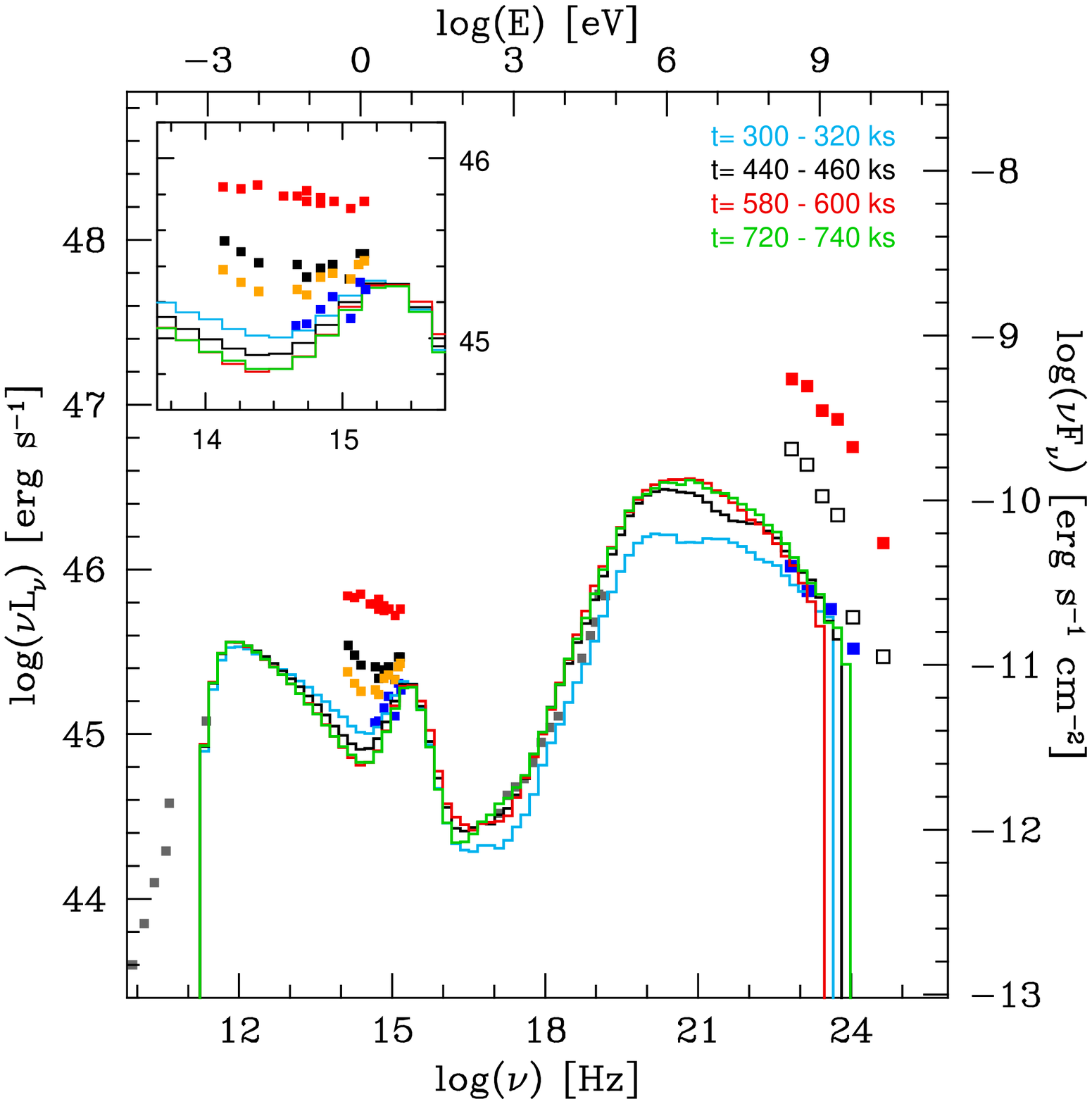}
}
\caption{
\textbf{Left, middle:} 
Light curves and SEDs for the quiescent state of the EC/BLR model, using a
blackbody approximation for the BLR spectrum.
The histograms in both figures are the results of our simulation, with the
energy band chosen shown in the legend of the left figure. 
The SED snapshots times are shown in legend of figures and marked 
with matching color segments in the light curves plot.
In this and all the paper SED plots the data points are multiwavelength SEDs of
PKS~1510$-$089 mostly in the spring of 2009 from
\citet{abdo_etal:2010:pks1510_multiwavelength_flares_2008_2009} and 
\citet{dammando_etal:2011:pks1510_with_agile_march_2009}.
The optical/infrared points, also show in the inset, are for the following dates:
blue for March 10, 
orange for March 18, 
black for March 19,
red for Match 25/26 (flare peak).
The \gray, \fermi/LAT spectra are:
blue squares for the quiescent state of the early 2009 
\citep[see][]{abdo_etal:2010:pks1510_multiwavelength_flares_2008_2009},
red squares the end of March 2009 flare,
black empty squares an earlier weaker flare (flare `a' in 
\citealp{abdo_etal:2010:pks1510_multiwavelength_flares_2008_2009}) which 
we show as a plausible reference for the gamma-ray state before and after
the March 2009 flare.
The gray points in radio, submm and \xray are not strictly simultaneous but we
regard them as representative because of the very modest variability exhibited
by the source during those months in these bands.
The \xray data include the XRT data averaged during the March 2009 flare, and
five year integrated BAT data in hard \xrays. 
\textbf{Right:} 
The SEDs for the quiescent state of the BLR model using the
\citet{tavecchio_gg:2008:blr_and_blazars} BLR spectrum (parameters are listed
in Table~\ref{tab:model_parameters}, \texttt{nf/blr}).
\label{fig:blr15nf_tave}
\label{fig:blr15nf_bb}
}
\end{figure*}
%-----------------------------------------------------------------------

%=======================================================================
\subsection{About model parameters freedom and constraints}
\label{sec:about_parameters}

In the EC models, we have 5 basic observables (variability time-scale
$\tau_\mathrm{var}$, synchrotron luminosity $L_\mathrm{sync}$,
estimated IC peak frequency $\nu_\mathrm{IC,p}$,
IC luminosity $L_\mathrm{IC}$, and \gray spectral index $\alpha_\gamma$) 
to constrain 6 free parameters ($R$, $B$, $n_\mathrm{e}$, $f_\mathrm{ext}$, 
$\gamma_\mathrm{min}$ and $p_\mathrm{inj}$). 
However, there are in fact additional constraints available to further limit the 
usable range of parameter space.
For example, as we will discuss later, in the EC cases the SSC flux can not be
too high in order to be consistent with the moderate \xray variability, which
in turn translates into a requirement on the magnetic field strength to be
sufficiently large.
On the other hand, the required diffuse stochastic acceleration time-scale 
($t'_\mathrm{acc}$) and hence the cooling time-scale can not be too short,
which then imposes a limit on how large the magnetic field can be.

Moreover, as already noted, the value of $p$ and $\gamma_\mathrm{max}$ in
the quiescent state are not direct input parameters. 
They are the results of the combination of $t'_\mathrm{acc}$ and
$t'_\mathrm{esc}$ (and the relevant cooling time-scale). 
The synchrotron peak frequency $\nu_\mathrm{sync,p}$ is not always used as a
constraint because for the physical conditions of our simulations it is not a
result of the electron distribution, but it is often determined by synchrotron
self-absorption.
At the same time, observationally its position in the SED is only poorly
constrained by currently available data.

For some other parameters, such as the bulk Lorentz factor of the jet, the
radius of the BLR or torus regions, values are set on the basis of empirical
estimates obtained by independent studies and therefore may not be regarded 
as truly free parameters.
On the other hand, because the energy density of the external radiation in the
blob frame, $U'_\mathrm{ext} \sim f_\mathrm{ext}\,\Gamma^2/R_\mathrm{ext}^2$, 
and $L_\mathrm{EC}/L_\mathrm{sync} \sim U'_\mathrm{ext}/B^2$, a given SED or
set of SEDs will impose a relationship between these parameters.

The size-scales of BLR and torus is generally found to follow a relationship like
$R \sim L_\mathrm{disc}^{1/2}$, with some range in the normalization typically
yielding a $R_\mathrm{BLR}= \text{few} \times 10^{17}$~cm and 
$R_\mathrm{IR}= \text{few} \times 10^{18}$~cm for a disc luminosity of the
order of that inferred in PKS~1510$-$089 on the basis of the observed BBB.
We thus decided to set $R_\mathrm{ext}$ to the value closed to the ones
expected from these estimates and given the uncertainty on the
$f_\mathrm{ext}$ to regard its value as a free parameter controlling the
normalization of the EC emission. 

%%%%%%%%%%%%%%%%%%%%%%%%%%%%%%%%%%%%%%%%%%%%%%%%%%%%%%%%%%%%%%%%%%%%%%%%%%%%%%%%
\vspace{-12pt}
\section{Results}
\label{sec:results}

As illustrated in Section~\ref{sec:pks1510} our aim is to reproduce the
quiescent and flaring states of PKS~1510$-$089.

We model a flare as being caused by an injection of relativistic electrons
ascribed to the effect of the interaction of the blob with a standing shock, 
as described in Section~\ref{sec:setup}.
As the shock travels through the blob (in the blob frame) it injects new
particles locally, \ie in the zones where it currently resides.
These newly injected electrons are treated in the same way as all other
electrons in each zone; they radiate through synchrotron and IC, and evolve
according to the same Fokker-Planck equation. 

We will review cases with BLR or torus as the dominant sources of external
photons for EC, as well as a pure SSC model, and compare light curves and SEDs 
with those of the March 2009 observations of PKS~1510$-$089.
In discussing the comparison we will mainly focus on the SEDs, in particular 
on the infrared, \xray and \gray bands.
The light curves, which we show for several observable bands, provided an
important constraint guiding the analysis and identifying suitable parameter
values but they do not illustrate the differences between different cases as
clearly as the comparison of data and SEDs for multiple epochs.  
This is also due to the fact that purely in terms of intensity variation in
narrow bands, the longest time-scale dominates (modulates) the time profile of
a flare, and in the cases discussed here the source crossing time is larger
than time-scales for electron processes. 
This was not true for instance for Mrk~421, for which as shown in
\citet{chen_etal:2011:multizone_code_mrk421} the light curve profiles 
for various \xray and TeV bands were different depending on whether the
particle related time-scales were faster or slower than the source crossing
time, providing us with additional diagnostics.

%-----------------------------------------------------------------------
\begin{figure*}
\centerline{
\includegraphics[width=0.49\linewidth]{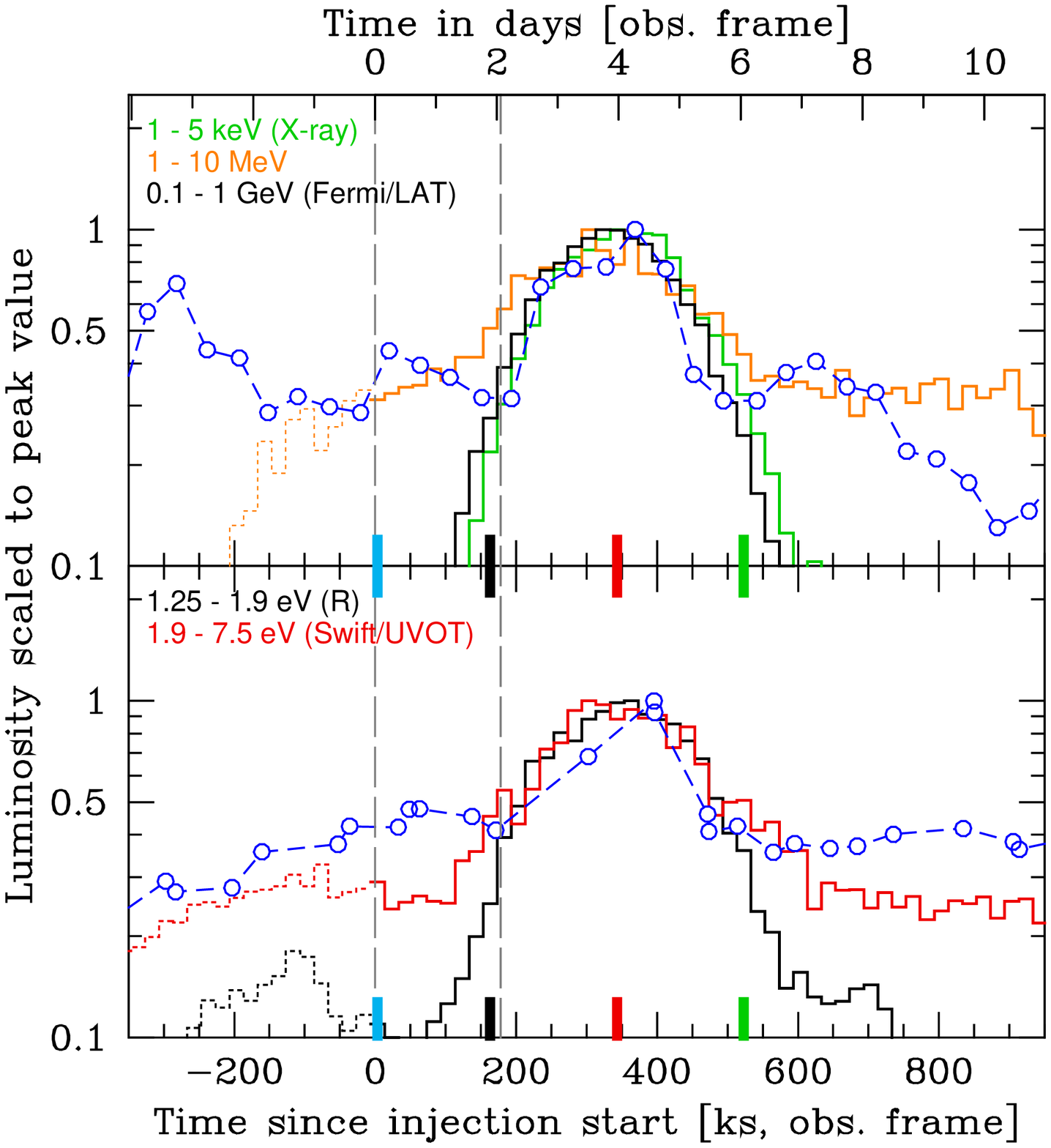}
\includegraphics[width=0.49\linewidth]{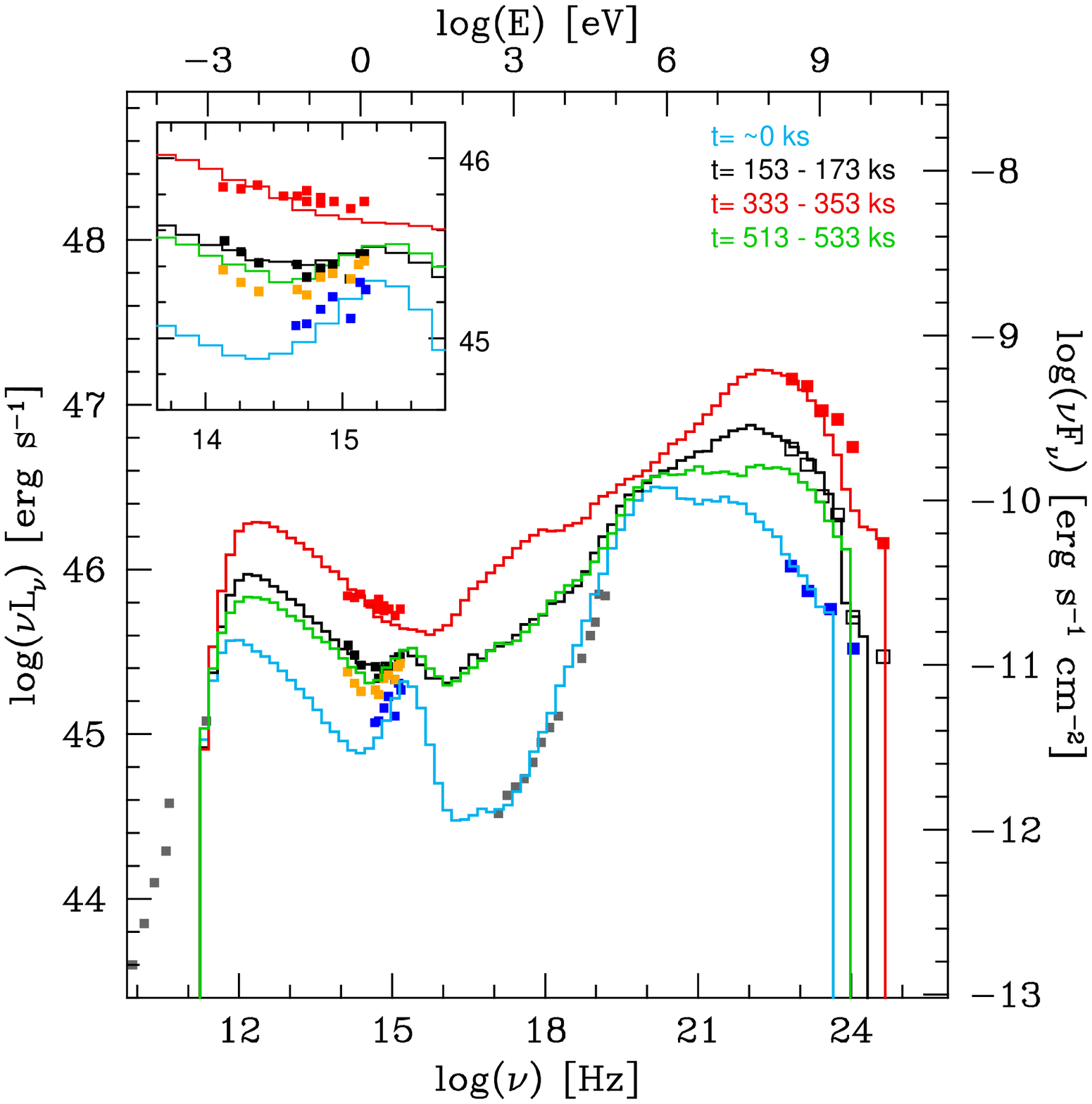}
}
\caption{
Light curves (left) and SEDs (right) for the first EC/BLR scenario. 
In these and all following light curve plots the blue circles in the lower
panel show the observed R-band light curve in March 2009 from the GLAST-AGILE
Support Program (GASP), those in the upper panel show the \fermi/LAT light curve
above 0.2~GeV in March 2009, simultaneous with the R-band data 
\citep{abdo_etal:2010:pks1510_multiwavelength_flares_2008_2009}. 
The SED data points are the same as those in Fig.~\ref{fig:blr15nf_bb}.
Simulation light curves are dotted before $T=0$ to emphasize that that interval
should be regarded as a setup time for the simulated region.
The two grey vertical dashed lines mark the start and end of the injection time,
\ie the time during which the shock is crossing the blob. 
The SEDs correspond to the times (in the observer frame) given in the legend,
and marked with colored ticks in the light curves plot.
Model parameters are given in Table~\ref{tab:model_parameters}, \texttt{blr15}.
\label{fig:blr15}
}
\end{figure*}
%-----------------------------------------------------------------------

%=======================================================================
\subsection{Quiescent state: no shock}
\label{sec:quiescent}

First we try to reproduce the quiescent state SED using as reference the period
of the last three months of 2008 during which the \gray flux measured by
\fermi/LAT remained low and with small variations (the `quiescent' period in
the naming adopted in \citealp{abdo_etal:2010:pks1510_multiwavelength_flares_2008_2009}).
We begin with using the blackbody approximation for the spectrum of 
the BLR emission. We show the results of this simulation in Fig.
\ref{fig:blr15nf_bb}, with the parameters used listed in Table
\ref{tab:model_parameters}.

In this simulation particle injection is exclusively given by the steady-state
pick-up term $Q_\mathrm{pick}$. 
Because there is no flaring activity, the flux level at every wavelength
reaches the steady state after a few light crossing times, and the light curves
remain almost flat except for statistical fluctuations. 
  
It is interesting to note that the R-band light curve reaches a flux level
higher than the quiescent level before it reaches steady state (effect
noticeable also in the SED in the infrared band). 
This is because the external photons need some time (of the order of one light
crossing time) to diffuse through the whole blob\footnote{%%%
The same is true for the internally produced synchrotron photons scattered
by SSC.  Their contribution to the photon field in each location will need a
time of the order of the light crossing time to stabilize, or, in the case
of variable synchrotron emission, to respond to the changes of intensity
happening elsewhere in the blob.}. 
During that time some zones which have not yet received the external photons
will experience significantly less IC cooling (which is dominant over
synchrotron cooling in this case). 
Hence the higher energy portion of their electron spectrum will remain at a
relatively higher level and produce a relatively more intense synchrotron
radiation during this phase.  
This is an example of how the light travel time effect can affect the actual
physics in the jet, not only just the way we perceive the emission.
While in this particular case the initial phase of the simulation and particle
evolution is not of astrophysical interest, this effect is realistic and
illustrates one important aspect of taking into account the finite size
of the source and the effect of light travel time on the physical evolution
of the system and its emission.

The SEDs match the low state data points at optical and \gray frequencies
pretty well. 
The radio data points do not match because it is likely that the radio emission
comes from additional emission regions rather than just the one producing the
optical and \gray emission.
The simulated spectrum in the \xray regime is much harder than the observed one. 
This improves significantly by using a more detailed description of the BLR
spectrum such as that discussed by \citet{tavecchio_gg:2008:blr_and_blazars},
as shown by the SEDs in Fig.~\ref{fig:blr15nf_tave}, where we used a BLR
spectrum obtained from their analysis.
This confirms their conclusion that an accurate treatment of the BLR spectrum
used for the EC emission is necessary in producing the spectrum in the soft to
medium \xray band ($\sim$0.1--10~keV), where usually high quality data are
available.  This band is of critical importance because it can provide
constraints on the relative contribution of SSC and EC, and also on the
characteristics of the lower energy end of the electron distribution.

%=======================================================================
\subsection{EC/BLR: shock crossing with $\mathbf{\Gamma=15}$}
\label{sec:blr15}

Starting from a baseline quiescent state like the one just discussed, we model
the flare for a case in which the dominant EC contribution is from the BLR.
For this first case we adopt a blob bulk Lorentz factor $\Gamma=15$.
The results are shown in Fig.~\ref{fig:blr15}.
Light curves show that the optical/infrared variability is larger in the R-band than 
in the bluer \swift/UVOT bands, due to the contribution of the non-variable big
blue bump emission in the UV.  This is consistent with observations. 

Both the \xray light curve and the SEDs clearly show that our simulations 
predict large-amplitude variations in the \xray flux. 
This is at odds with observations which show only modest \xray variability
(less than a factor of 2) throughout the entire 2009 observing campaigns,
including the largest flares.
The excessive \xray variation in our simulations is the result of SSC emission.   
Although currently our model does not track separately photons of EC and SSC
origin, the parameters we use indicate that emission by SSC is not negligible
in the \xray band even in the quiescent state.  
In case of a flare caused by changes in the electron spectrum and/or density
the amplitude of variation of the SSC emission will always be larger than that
of the synchrotron (as long as we are considering electrons scattering in the
Thomson regime, which is the case here).  
Therefore as we model the factor of 10 increase of the non-thermal emission in
the optical/infrared band the corresponding IC emission will vary by a factor
up to a 100, with a large contribution in the \xray band.

This SSC variation makes this model and parameter set not consistent with
the observed features of PKS~1510$-$089. 
This example also illustrates the importance of modeling the time evolution of
the SED, rather than simply modeling with sets of parameters left fully free to
vary between different epochs the high and low state SEDs, or even sequences of
SEDs taken close enough in time during a single flare that they are very likely 
to be related to each other as part of the development of a single event.

%-----------------------------------------------------------------------
\begin{figure*}
\centerline{
\includegraphics[width=0.49\linewidth]{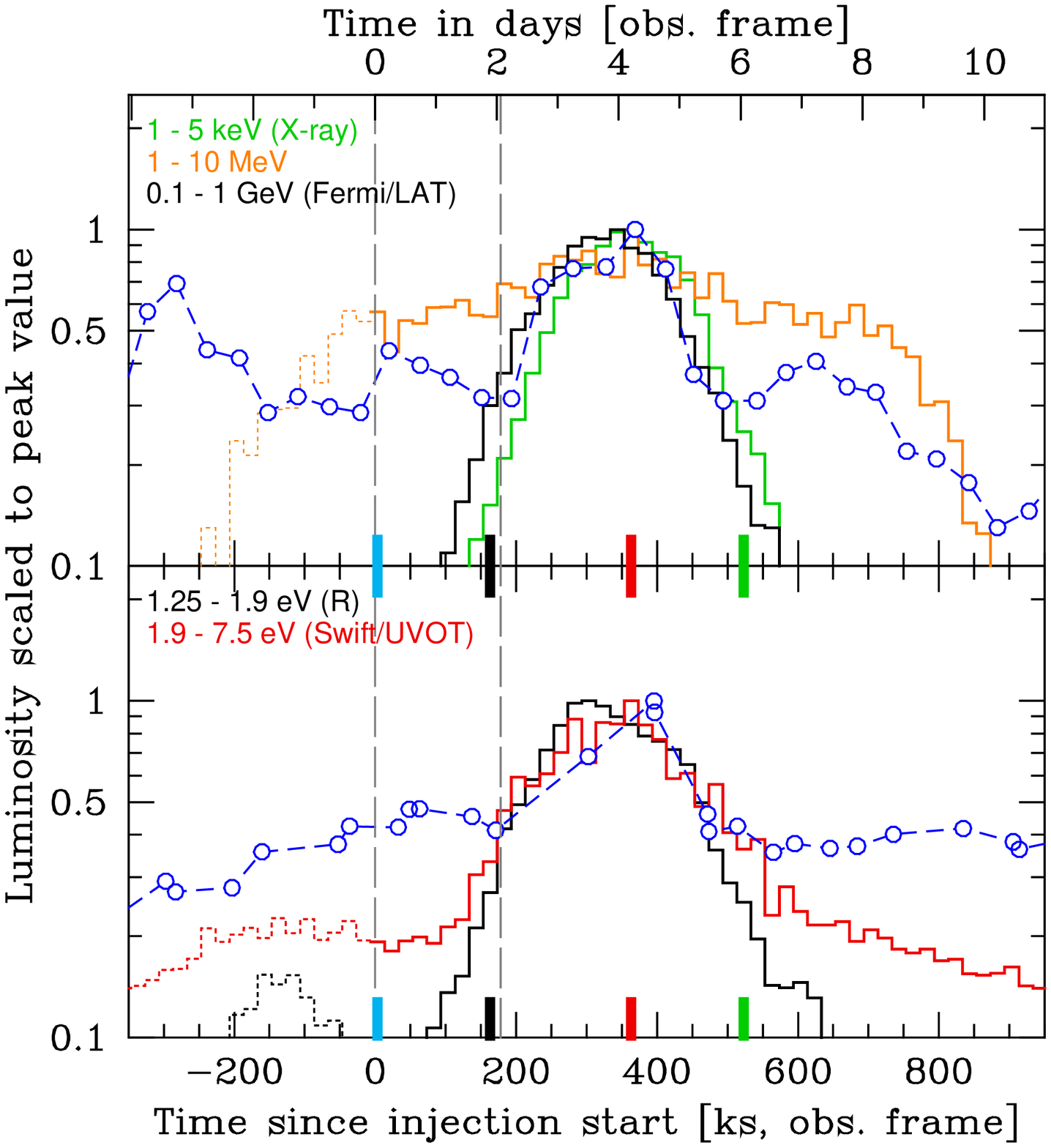}
\includegraphics[width=0.49\linewidth]{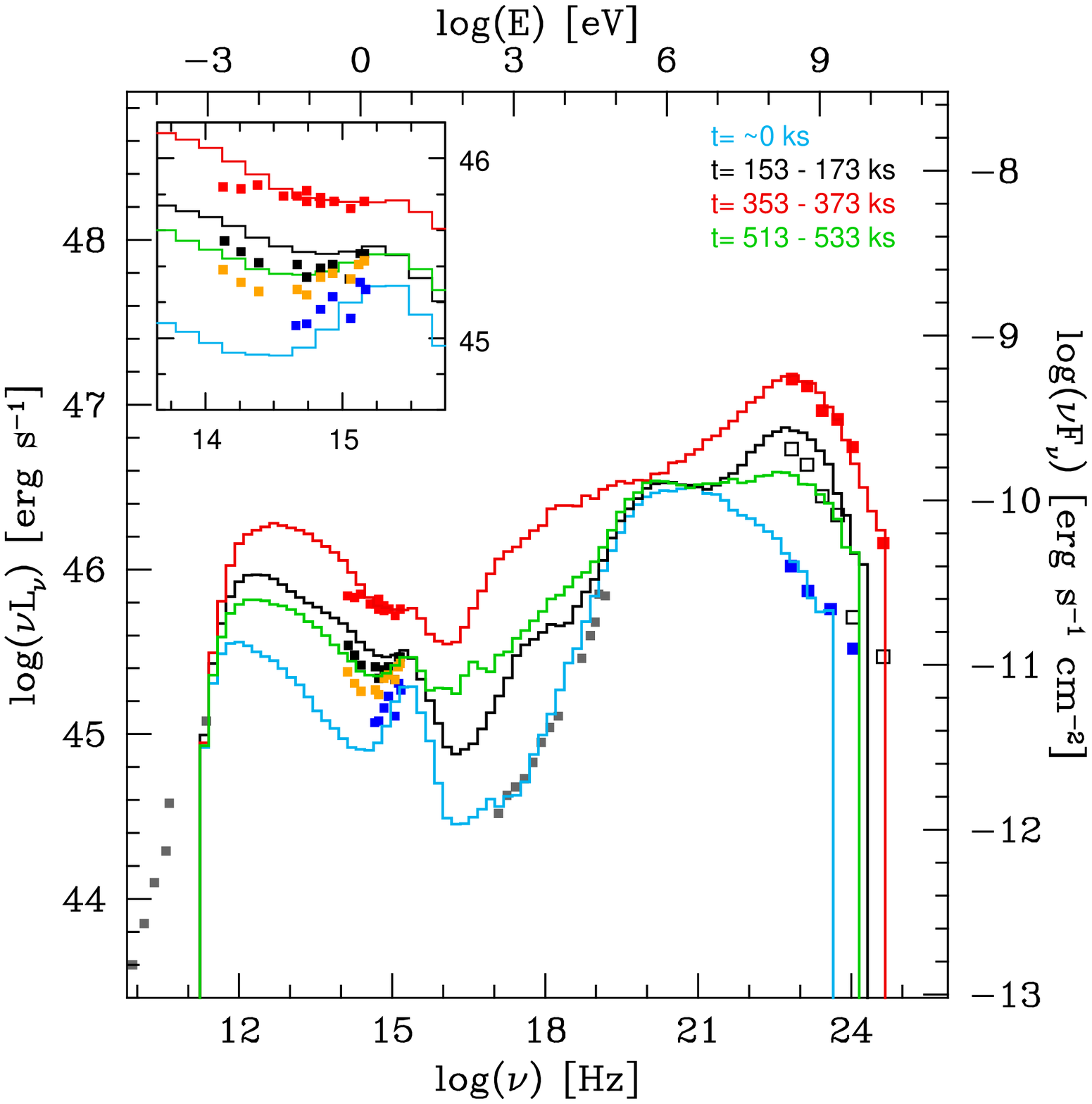}
}
\caption{
Same of Figure~\ref{fig:blr15} for the second case of EC/BLR flare simulations, 
with $\Gamma=15$ and higher minimum energy for the injected electrons,
$\gamma_\mathrm{min}=90$ (see Table~\ref{tab:model_parameters}, \texttt{blr15highgmin}).
\label{fig:blr15_highgmin}
}
\end{figure*}
%-----------------------------------------------------------------------

\subsubsection{Quiescent vs. flaring state and the importance of time-dependent
multi-zone modeling}

The comparison between the SEDs of the steady and flaring states shown
in Figures~\ref{fig:blr15nf_tave} and~\ref{fig:blr15} illustrates another
important difference between a time-dependent multi-zone simulation and 
a one-zone (effectively point-like) simulation, for which the quiescent case
can be considered a proxy\footnote{The steady state emission is effectively
equivalent to what would be produced by a one-zone non-time-dependent code.}: 
the SEDs of the more realistic model are significantly more complex in the
shape and variation of the high energy component.  
This is the result of both internal and external LTTE giving to the observer a
mix of emission produced at different times in zones at different stages of
the flare development and with electron distributions at different stages of
their evolution.

Even if locally the processes affecting the electrons are fast and the particle
spectrum could be regarded as reaching rapidly a steady state (in case of
injection lasting for a long enough time), the sequence of SEDs produced in a
flare is not equivalent to a sequence of steady state SEDs \citep[see][for an
example of this approach applied to the FSRQ 3C~454.3]{bonnoli_etal:2010:3c454_brightest_days}.

Admittedly in this paper we are presenting one possible scenario for the flaring
state of a FSRQ, but this type of differences can be expected to exist for a 
wide range of plausible scenarios of variable emission from a relativistic jet.

%-----------------------------------------------------------------------
\begin{figure*}
\centerline{
\includegraphics[width=0.49\linewidth]{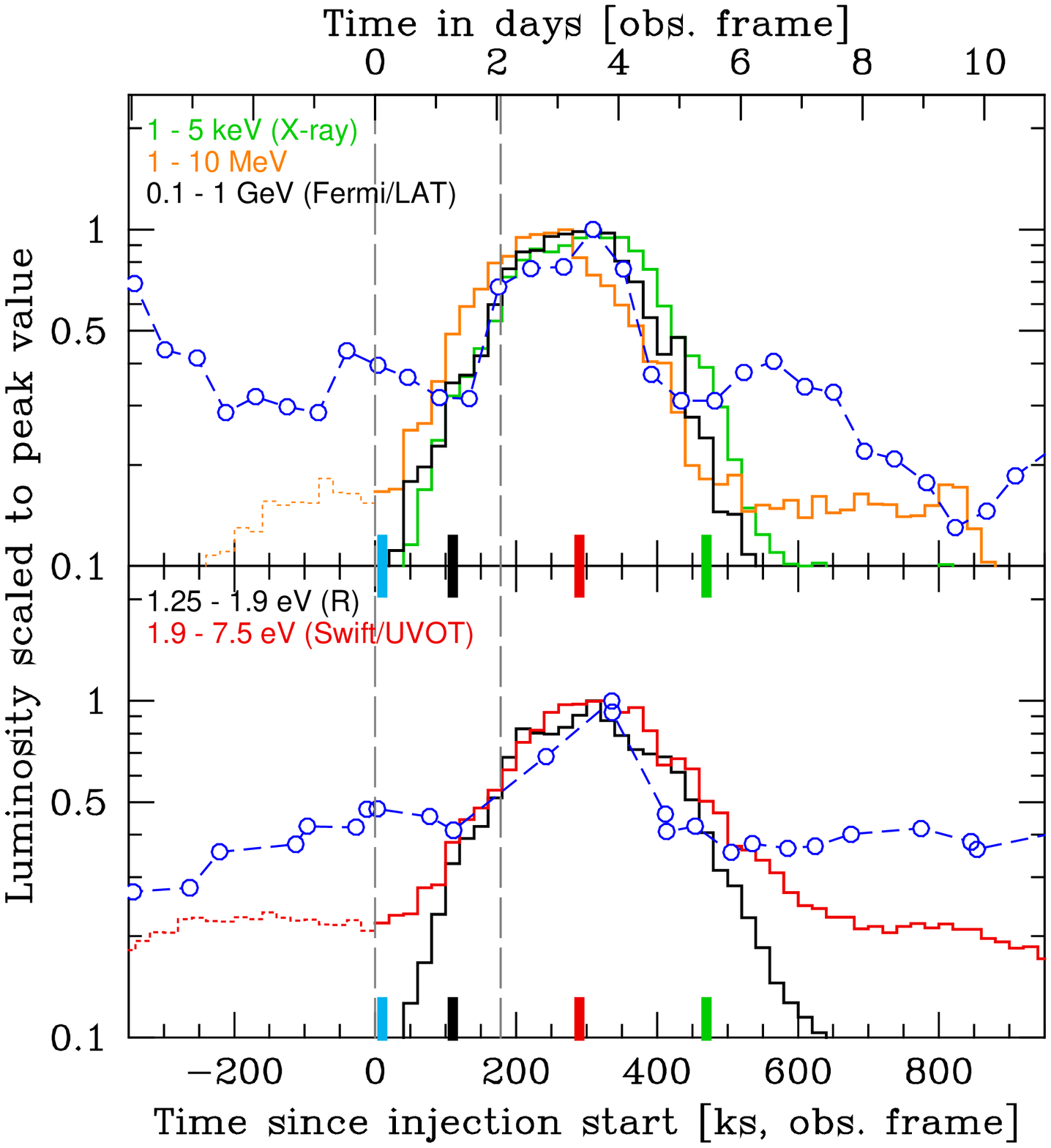}
\includegraphics[width=0.49\linewidth]{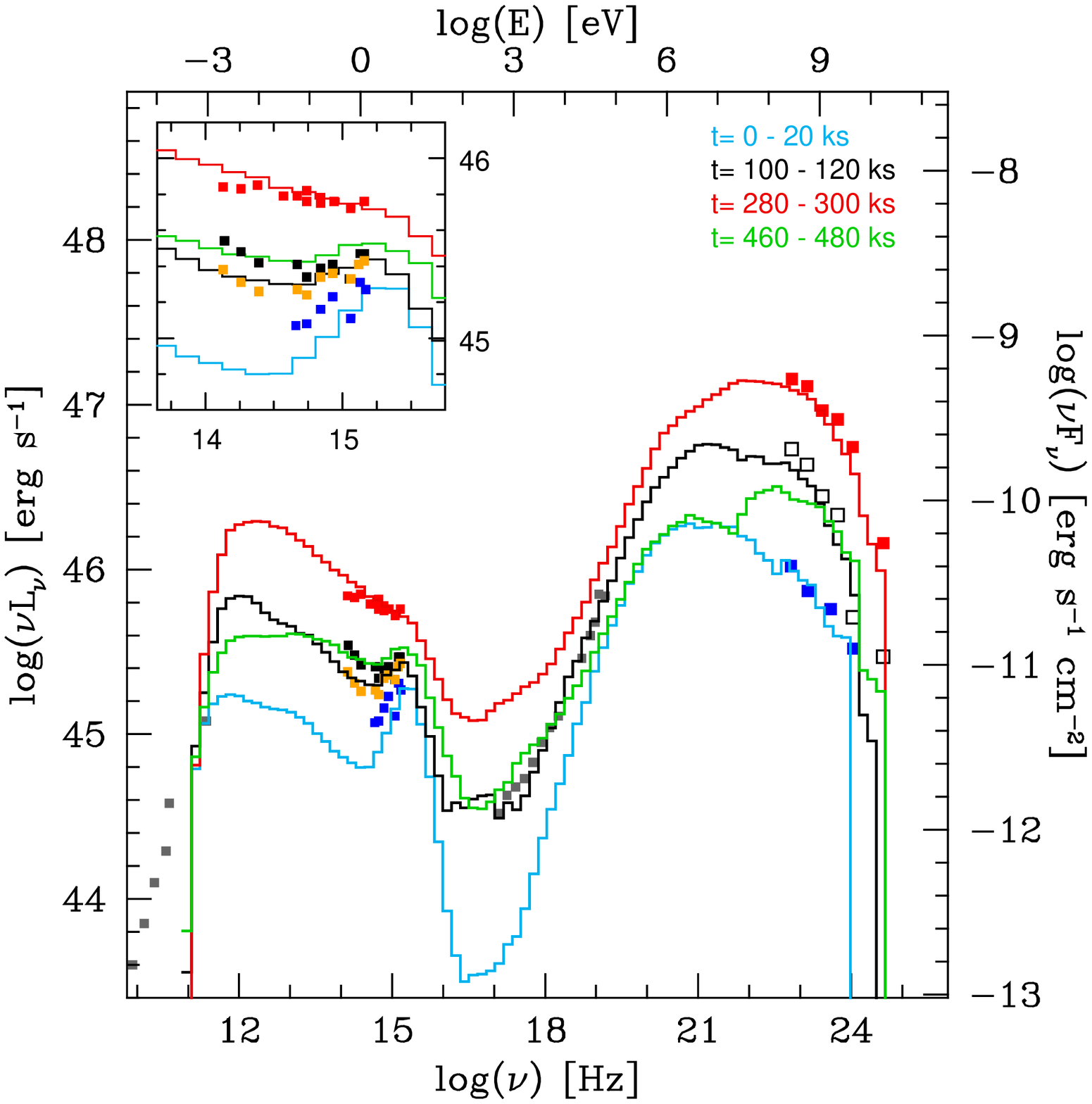}
}
\caption{
Light curves (left) and SEDs (right) for the EC/BLR case with higher bulk
Lorentz factor, $\Gamma=25$.
Parameters are given in Table~\ref{tab:model_parameters}, \texttt{blr25}.
\label{fig:blr25}
}
\end{figure*}
%-----------------------------------------------------------------------

%=======================================================================
\subsection{EC/BLR: $\mathbf{\Gamma=15}$, with higher $\mathbf{\gamma_\mathrm{min}}$}
\label{sec:blr15_highgmin}

A potential remedy for the excessive \xray variability crisis could be to
increase the minimum energy of the injected electrons.
The peak of the emission of the variable component, driven by the electron
injection, is determined by the electron's $\gamma_\mathrm{min}$.
A higher $\gamma_\mathrm{min}$ may push the SSC emission by these electrons
to peak at a higher frequency where it can be `hidden' beneath the
rapidly rising stronger EC emission. 
However, $\gamma_\mathrm{min}$ is fairly constrained by the observed MeV-GeV
\gray spectrum, namely by the fact that we do not observe a spectral turn over
in the \fermi/LAT data.  
Too large a $\gamma_\mathrm{min}$ will produce a EC SED peaking in the \fermi/LAT 
observational band. 
With this in mind, we tested one case with a slightly higher injected
$\gamma_\mathrm{min}=90$. 
The results are shown in Fig.~\ref{fig:blr15_highgmin}.

The peak of the EC SED in this case is shifted close to the lowest energy data
point of the \fermi/LAT spectrum, indicating that $\gamma_\mathrm{min}=90$ is
already of the order of the largest value that we can use with the current BLR
spectrum and a Lorentz factor of 15.
On the other hand, while there is some change on the spectral shape of the
high state \xray spectra, the fundamental problem of the excessive amplitude of
its variation is not mitigated. 
The inconsistency between the observed and simulated \xray variability remains
a problem in the higher injected $\gamma_\mathrm{min}$ case.

%=======================================================================
\subsection{EC/BLR: higher Lorentz Factor, $\mathbf{\Gamma=25}$}
\label{sec:blr25}

Instead of moving the SSC emission in frequency, another route to solve the
\xray variability inconsistency may be to decrease the level of the SSC
emission relative to the other components.
In order to do this while keeping the same level of synchrotron emission, we
need to decrease the ratio between the synchrotron (SSC seed photons) energy
density and the magnetic energy density. 
We can achieve this by increasing the Doppler factor because in that case the
synchrotron energy density in the blob frame needs to decrease accordingly to
match the optical data.  
This requires to change the values of other parameters to produce a SED well
fitting the observed one.  The modified parameters are reported in
Table~\ref{tab:model_parameters}.  The results are shown in Fig.~\ref{fig:blr25}.

In this case, the luminosity in the \xray dip between the two main components
is indeed lower.  Nevertheless because any SSC emission occurring in this band
would still be varying by a factor larger than that of the optical flux, the
range of the \xray variation remains large, and easily exceeding the constraint
set by the well measured \xray intensity and spectrum.
However, since in this case the quiescent \xray flux is lower than the observed
one, it may be possible to explain the \xray band spectrum as comprising a
contribution from additional, relatively cooled blobs, which would partially
dilute the large variation.
However, even taking that into account, considering the spectral variability
present in the varying \xray component, it may not be straightforward to
reconcile the overall \xray properties with the remarkably stable observed
spectra.  

%=======================================================================
\vspace{-12pt}
\subsection{EC dominated by IR emission from the torus}
\label{sec:torus15}

Emission from hot dusty, molecular, gas in the putative torus surrounding
the accretion disk is another plausible source of external photons in the
immediate environment of the relativistic jet.
This scenario is motivated by the observation of the coincidence of
\gray flares with the appearance of new knots in radio images of 
PKS~1510$-$089 \citep{marscher_etal:2010:pks1510}.
These observations hint that the emission region responsible for \gray flares
is located at parsec scales, which is beyond the usually inferred radius of the
BLR \citep{kaspi_etal:2007:reverberation_mapping_of_luminous_quasars,
bentz_etal:2006:Rblr_vs_L_relationship_for_AGNs,
bentz_etal:2009:Rblr_vs_L_relationship_for_AGNs_paper2}.
At this distance, the IR torus
\citep{pier_krolik:1992:radiation_supported_tori,pier_krolik:1992:torus_spectra} 
becomes the main candidate as the source of EC seed photons 
\citep{gg_tavecchio:2009:canonical_blazars,sikora_etal:2009:constraining_emission_models}.

\begin{figure*}
\centerline{
\includegraphics[width=0.49\linewidth]{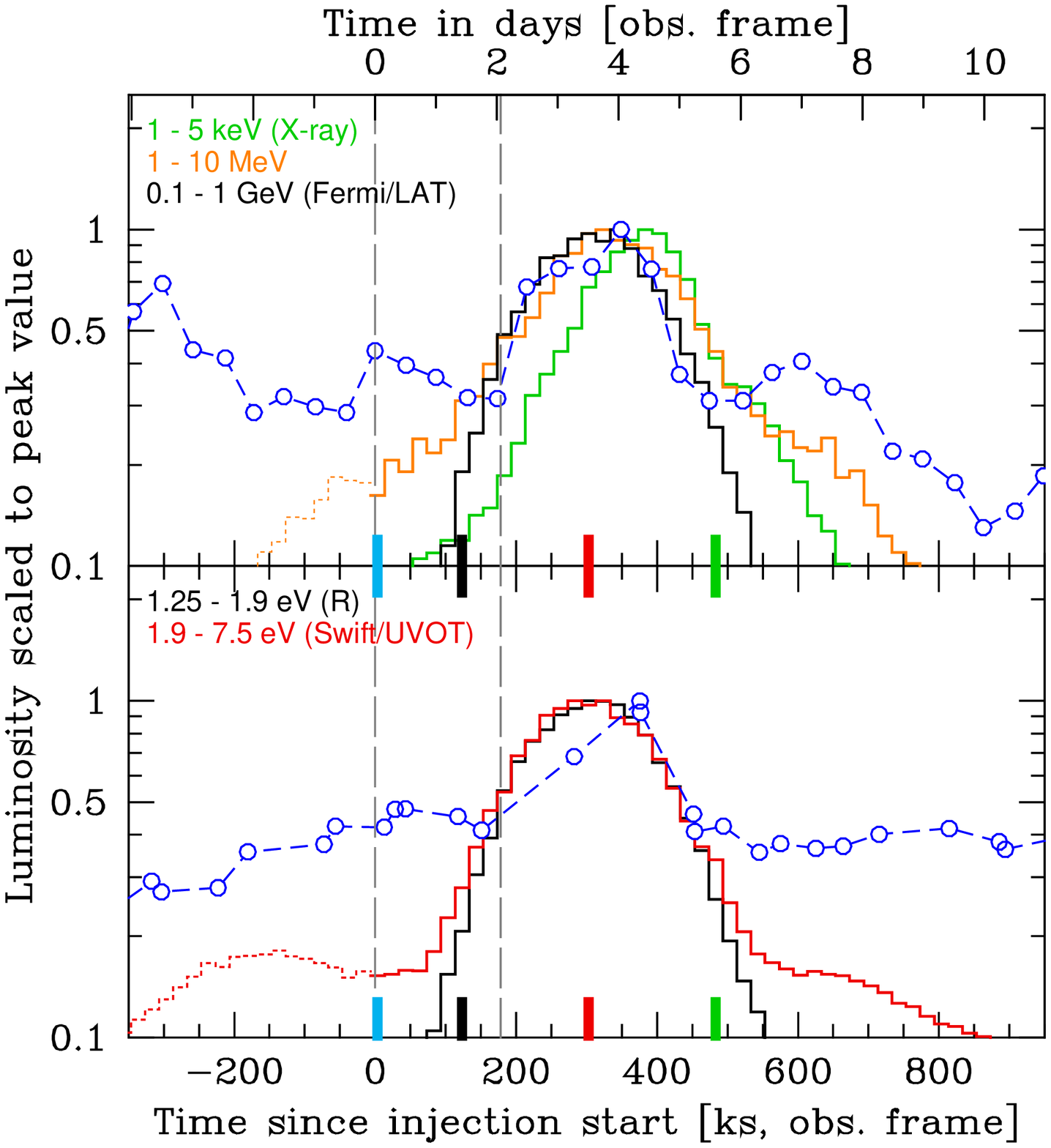}
\includegraphics[width=0.49\linewidth]{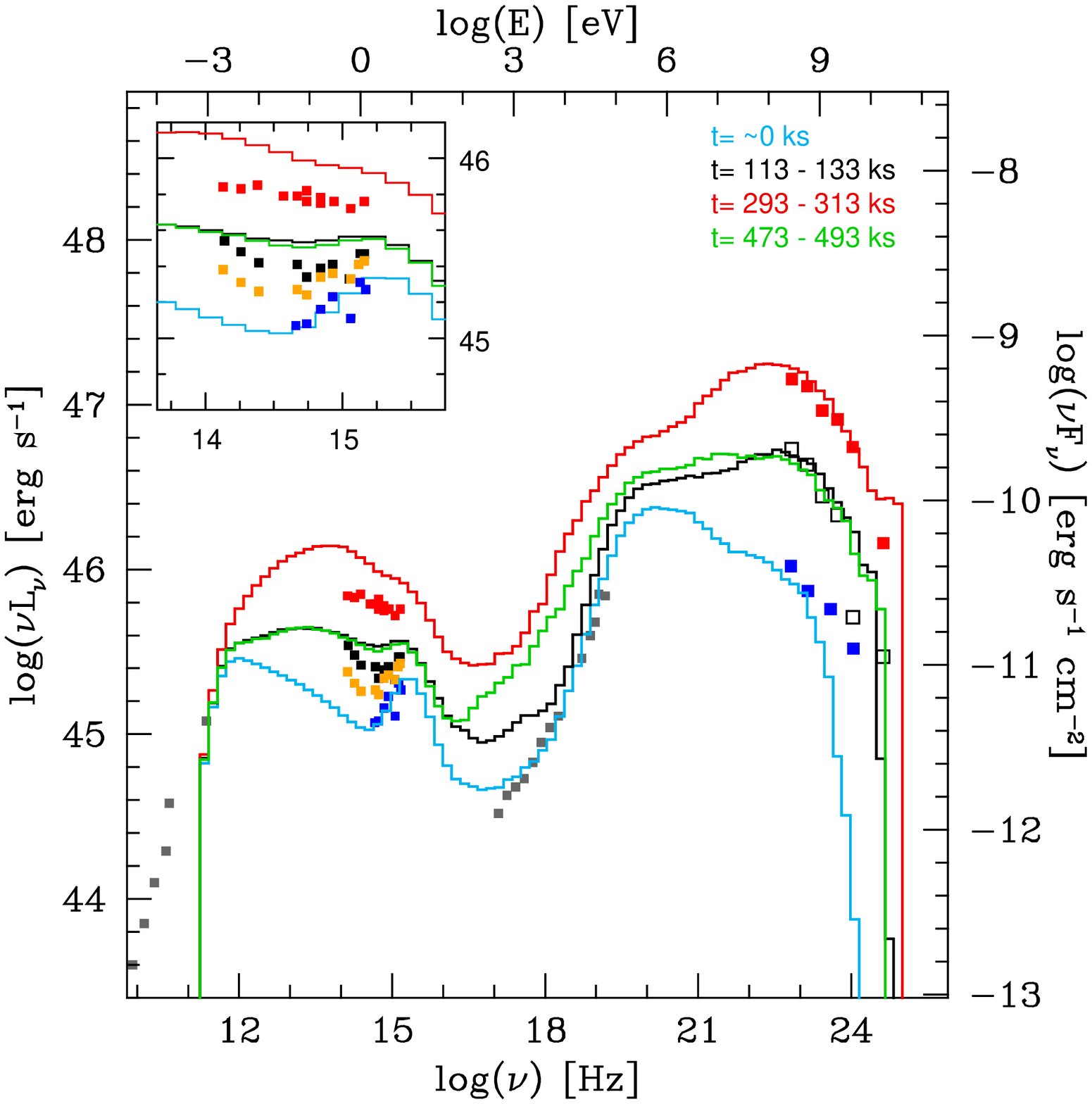}
}
\caption{
Light curves (left) and SEDs (right) for the case with the EC 
dominated by radiation from the torus.  Parameters are listed in
Table~\ref{tab:model_parameters}, \texttt{torus15}.
\label{fig:torus15}
}
\end{figure*}

We calculate the energy density and temperature of the torus emission according
to equations (\ref{eq:u_ir}) and (\ref{eq:T_ir}).  
Because the energy density and SEDs of the emission from the torus are very
different from those of the BLR, it is necessary to change the values of
several parameters to be able to match the quiescent and then flaring states.
The best set of parameters is reported in Table~\ref{tab:model_parameters}, and the 
light curves and SEDs in Fig.~\ref{fig:torus15}.

The broadband SEDs compare reasonably well with observations; the light curves
vary on a time-scale consistent with the data; the optical and \gray light
curves are well correlated with no significant lags; the variations in the
optical and \gray bands have similar amplitude; the variations in the UV band
are less prominent than those in the optical.
However, it is worth noting that it has proven to be difficult to concurrently
match the GeV spectrum and the soft slope through the infrared bands.
Although we can not claim to have achieved a perfect match in the latter in the
previous cases, this is an issue that did not seem to emerge as seriously as
here.

%-----------------------------------------------------------------------
%-----------------------------------------------------------------------

The problem of \xray variability in this case is similar to the one in the
BLR case.  It is likely that similarly to the previous case, a higher Doppler
factor would lower the flux produced by SSC in the \xray band.

One of the main differences between the torus emission and the BLR emission as
source of EC seed photons is that the owing to its lower temperature the torus
radiates at lower frequency compared to the BLR.  
This means that to scatter these seed photons to the same \gray energies, the
energy of the electrons needs to be higher than those needed in a BLR scenario.
In the context of the scenario presented in this work, this means that more
efficient particle acceleration is needed to sustain the high energy electrons,
which have faster cooling times.
This also means that faster particle escape is to be expected in the torus
scenario, otherwise the accelerated particles will not form a power-law
distribution that can produce emission with the observed spectral shape.
It turns out that the value for the particle escape time-scale parameter needed
in this case may be too fast to be realistic ($t'_\mathrm{esc}=0.015\, Z/c$).
However, it is worth emphasizing that, chosen name notwithstanding, the
`escape' term in the kinetic equation may be regarded as a crude approximation
for a describing a generic energy independent process affecting electrons, and
in this sense a value significantly smaller than the source crossing time may
not be automatically be considered unphysical.
Nevertheless, because the time-scale for plausible candidate such as actual
escape or adiabatic cooling processes would likely be of the order of $Z/c$, 
it would certainly be desirable to not be forced to such extreme values for
$t'_\mathrm{esc}$, and we regard this as a serious problem for EC on torus
emission, in the framework discussed here.

%=======================================================================
\subsection{Pure SSC}
\label{sec:ext_ssc}

We already cited some of the recent results suggesting that the active, \gray
emitting region, in the jets of several blazars may be located beyond the
size-scale of the BLR on the basis of correlated multiwavelength variability
and VLBA imaging 
\citep{larionov_etal:2008:multiwavelength_campaign_3C279,
sikora_moderski_madejski:2008:3C454_blazar_zone,
marscher_etal:2010:pks1510,
agudo_etal:2011:OJ87,
agudo_etal:2011:0235}.
Additionally, the detection in TeV \grays of a few FSRQs, including a recent
confirmation for PKS~1510$-$089 \citep{atel_3965:2012:pks1510_magic_detection}
(others are 3C~279, \citealp{albert_etal:2008:3C279_TeV_detection,
aleksic_etal:2011:MAGIC_3c279}, and 
PKS~1222$+$216, \citealp{aleksic_etal:2011:MAGIC_pks1222}), 
challenges traditional EC scenarios because jet emitted TeV photons would be
readily lost by photon-photon pair production with the copious soft photons
surrounding the jet. 
These findings have thus stimulated a renewed interest in the possibility that
even for some of these powerful jets in systems with luminous accretion disk
and broad line emission, the \gray emission may be predominantly by SSC
\citep[for analysis on 3C~279 see \eg,][]{boettcher_reimer_marscher:2009:VHE_3c279}.
Finally, studying a large sample of well characterized blazars detected 
by \fermi/LAT and with an estimate of their jet intrinsic power,
\citet{meyer_etal:2012:EC} find that, while for the highest jet power objects
there is a clear collective sign of EC being the dominant mechanism for their
\gray emission, the properties of the rest of the population are consistent
with SSC.

%-----------------------------------------------------------------------
\begin{figure*}
\centerline{
\includegraphics[width=0.49\linewidth]{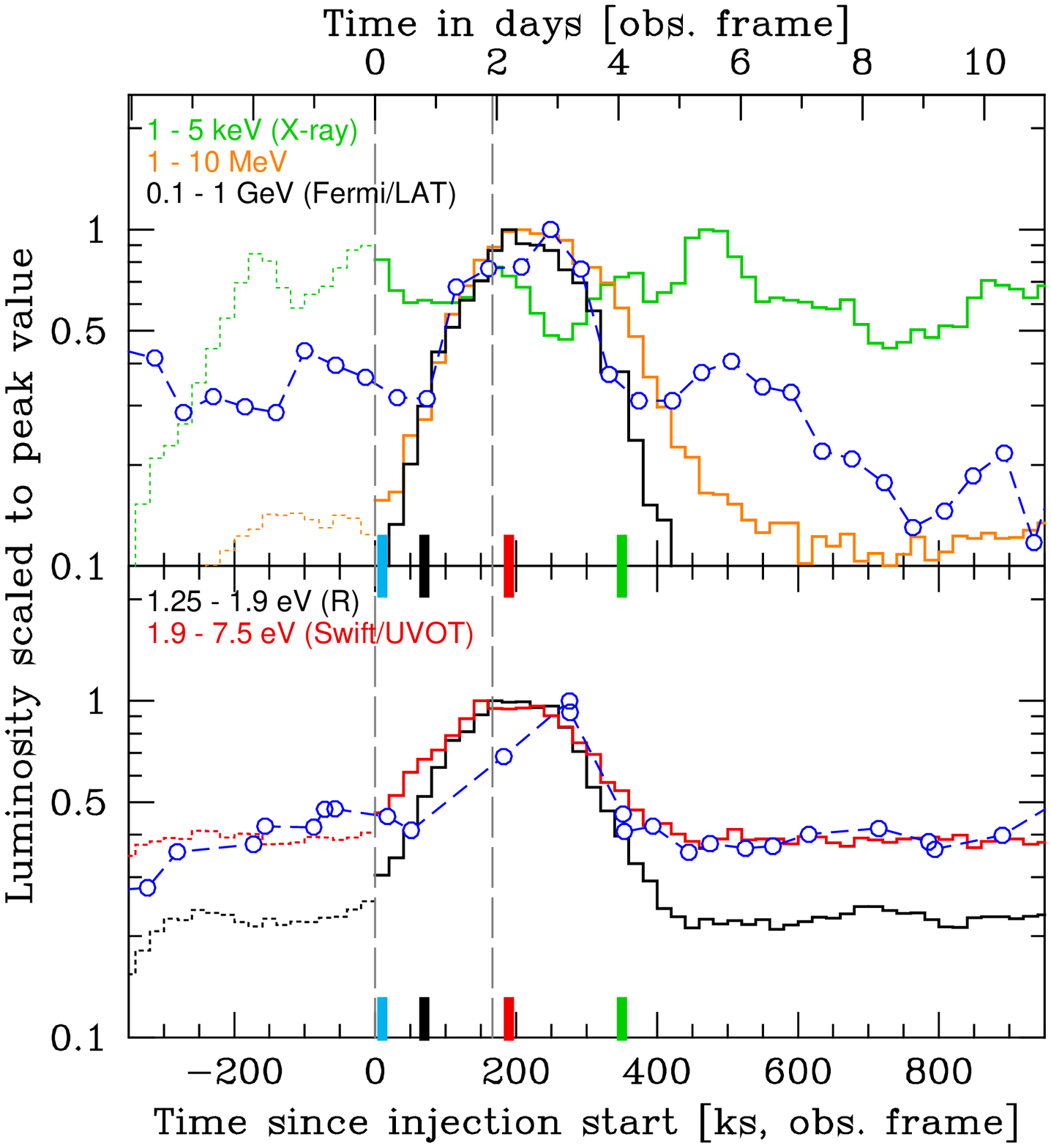}
\includegraphics[width=0.49\linewidth]{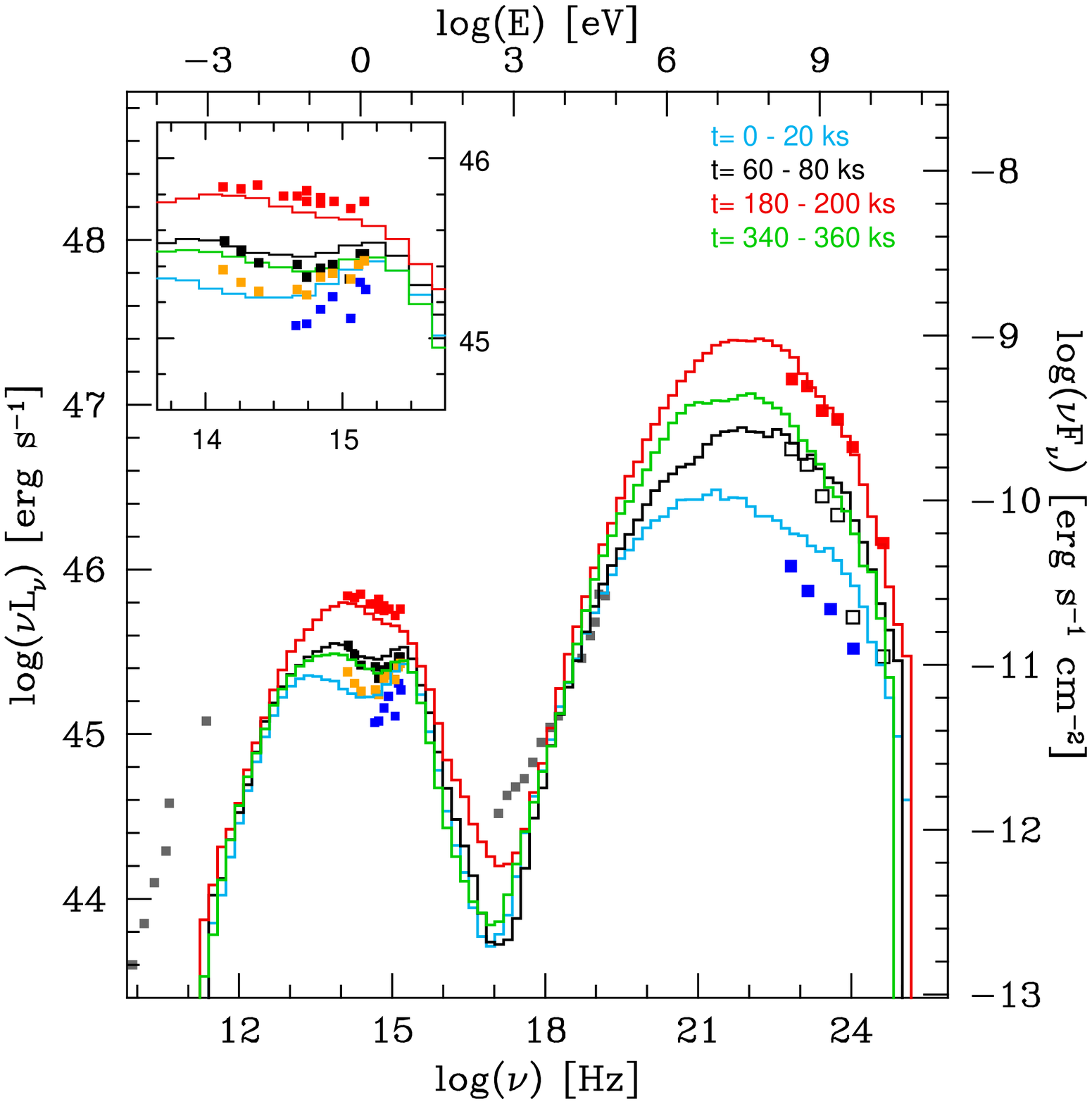}
}
\caption{
Light curves (left) and SEDs (right) for the SSC case.
Parameters are listed in Table~\ref{tab:model_parameters}, \texttt{ssc}.
\label{fig:ext_ssc}
}
\end{figure*}
%-----------------------------------------------------------------------

Moreover, as discussed in the previous sections, our simulations of EC models
for the type of flaring blob scenario presented here show large variability in
\xray which is inconsistent with the very robust observational finding of lack
of significant variations even during the flaring phase, and which is caused by
the SSC contribution.
The existence of this latter is unavoidable, and our tests suggest that even
mitigating its effect is not a trivial endeavor (\S~\ref{sec:blr15_highgmin},
\S~\ref{sec:blr25}).  

Therefore, it seems natural to want to test if we could reproduce the benchmark
observations with a pure SSC scenario.
The results are shown in Fig.~\ref{fig:ext_ssc}.

The SSC model reproduces well several aspects of the observations.
In particular, the amplitude of the flux variability and the spectral change 
in the \xray band are overall much closer to what was observed.
The \fermi/LAT band spectrum and intensity is also well matched.
Because of the presence of the steady disk emission mostly contributing to the
blue/UV flux, the flux in the R-band varies less than the \swift/UVOT band. 

However, the simulations produce an \xray spectrum consistently harder than
the one observed, mostly determined by the sharp `cut-off' of the lower energy
end of the electron distribution.  
The observed SED, namely the very hard \xray spectrum, constrains this latter
to a fairly high $\gamma_\mathrm{min}$, much larger than the values used in the
EC cases.
The simulated infrared spectrum is also somewhat harder than the spectrum
observed in the intermediate state.
Finally, the frequency at which the synchrotron spectrum peaks tends to be too
high, which is a consequence of the high $\gamma_\mathrm{min}$ required by the
\xray spectrum.
Since the SSC model has fewer free parameters than the EC model, they are more
constrained than those in the EC models, and do not leave us much freedom for
improving significantly on these issues.

%%%%%%%%%%%%%%%%%%%%%%%%%%%%%%%%%%%%%%%%%%%%%%%%%%%%%%%%%%%%%%%%%%%%%%%%%%%%%%%%
\section{Discussion and Conclusions}
\label{sec:discussion}

We have modeled multiwavelength variability produced in the jet of
PKS~1510$-$089, with EC model involving external radiation from BLR or
dusty molecular torus, as well as pure SSC model. 

As discussed also by \citet{tavecchio_gg:2008:blr_and_blazars}, the use of a 
realistic BLR spectrum seems indeed to be critical to model accurately the 
inverse Compton component, in particular its lower energy end which is observed
in great detail in the \xray band, thus providing a powerful diagnostic 
on model parameters. 
The results presented here, namely the challenging issue of the large \xray
variability caused by the SSC contribution, which has a distinctly different
spectral shape from the EC, further strengthen the importance of modeling
as accurately as possible the source of the EC seed photons.

In the same context, we should comment on the fact, noted in
Section~\ref{sec:about_parameters}, that in order to obtain reasonable SEDs and
light curves in the EC/BLR cases we had to adopt a reprocessed fraction of
BLR luminosity, $f_\mathrm{BLR}$, that is significantly smaller than what
one would expect. 
A more traditional $f_\mathrm{BLR}$, such as 0.1 in
\citet{gg_tavecchio:2009:canonical_blazars}, would yield a more intense BLR
radiation energy density and the increased electron energy loss rate in the
blob would push the model towards fairly extreme values for several of the
parameters. 
This could be regarded as an indication that the general scenario that we
studied in this work may not be a suitable explanation for flares in
FSRQs.  On the other hand, this scenario yields acceptable, though not
perfect, results for the EC/torus and the SSC cases.

As discussed in Section~\ref{sec:about_parameters}, the combination of
$(f_\mathrm{ext}\,\Gamma^2/R_\mathrm{ext}^2)/B^2$ is related to the ratio
between the IC and the synchrotron luminosities (Compton dominance), which is
an observable.  
Although its value may vary during a flare, the value of the Compton dominance 
imposes a relationship between those parameters of the source (hence model). 
A given source or source state may have a preferred value for the Compton
dominance which should be matched by a combination of $f_\mathrm{ext}$,
$\Gamma$ and $R_\mathrm{ext}$ (and $B$), independently on whether one wants to
model it with EC/BLR or EC/torus.
For the Compton dominance ratio that seems to work well for PKS~1510$-$089 it
is possible to adopt plausible values or $R_\mathrm{ext}$ and $f_\mathrm{ext}$
for the EC/torus case and less so for the EC/BLR case. 
This consideration does not depend strongly on the scenario adopted for the
the blob and the mechanism causing outbursts and it may suggest that indeed
PKS~1510$-$089 may not be modeled by EC/BLR in general.
Alternatively, it could mean that our naive idea of the BLR and torus does not
represent their real structure and causes the EC/BLR model to fail to match the
data. Another possibility is that, the relativistic jet in this source has a 
milder bulk Lorentz factor $\Gamma$ than 15 or 25 that we adopted in this paper.
However this in conflict with the high $\Gamma$ required to avoid the large SSC
contribution.

Leaving aside the above considerations, and looking at the results of the
simulations as run with the best fitting parameter sets, it is difficult
to identify a clear superiority for any of them.
None can be convincingly excluded.

The EC/BLR scenario suffers the problem of the large \xray variability, and it
produces better, marginally satisfactory, results if we adopt a larger value
for the bulk Lorentz factor, which is still within the observed range. 
It also requires parameters for the BLR which are at least unusual.

The EC/torus also has the problem of highly variable SSC in the \xray band,
which we can expect to be mitigated by using a larger Lorentz factor, and it 
works with reasonable parameters for the external radiation component, but 
it requires very fast particle acceleration and escape time-scales to
maintain the balance between acceleration and cooling for the high energy 
electrons necessary to scatter to the GeV band the lower energy external photons.
  
The pure SSC model naturally addresses the \xray variability issue, but at
price of very high minimum electron energy to avoid significant emission in
\xrays, which in turn make it difficult to reproduce the synchrotron peak
region SED and makes the \xray spectrum significantly harder than the observed one.
Because its GeV emission comes from scattering lower energy photons than the
EC/BLR case, like the EC/torus case the SSC requires fast escape time-scale,
though less extreme, which could be considered a serious problem.

In all three cases, their \xray problems may be relaxed if there is a slowly
varying contribution from other emission regions, filling the dip between the
synchrotron and IC components, softening the \xray spectrum and diluting its
variations.

Overall, while within the framework studied in this work we can not firmly
discriminate the three main scenarios, and the framework that we adopted is
just one possibility, our results show clearly the differences produced by a
more realistic treatment of the emitting source in the shape of SEDs and their
time variability over relevant, observable time-scales. 

These results demonstrate that proper modeling of the high quality data
produced by the plethora of best multiwavelength campaigns on the brightest
\fermi/LAT blazars to exploit the wealth of information that they carry and
advance our understanding of their physics can only be achieved with
time-dependent multi-zone simulations.

Looking forward, there are several aspects of the predicted source variability
that we have not discussed, such as more detailed analysis of correlated
variability (\eg time lags).
Moreover as shown here, SEDs exhibit complex features and variations and
it may be possible to explain the interesting, challenging, observation of
large break in the \fermi/LAT spectra seen in a handful of sources, comprising
both FSRQs and low-peaked BL\,Lacs \citep{abdo_etal:2010:spectral_properties_of_blazars}.
The observed breaks are larger than what would be produced by cooling and too
sharp to be consistent with an exponential cutoff.
Several ideas have been proposed to explaining them as due to external factors,
namely $\gamma\gamma$ absorption outside the jet \citep{poutanen_stern:2010:GeV_breaks,%
finke_dermer:2010:GeV_break_3c454,ackermann_etal:2010:3c454_fermi_outburst}, 
and some are somewhat inconsistent with other multiwavelength properties.

Finally, in these simulations we have used a simple model of energy independent
particle acceleration, whereby in order to maintain a power-law electron
distribution with a certain slope, the particle escape time-scale is tied to
the acceleration time-scale. 
The fast radiative cooling in our model requires fast particle acceleration and
hence fast particle escape.
This required escape time-scale has turned out to be smaller than the limit of
$Z/c$, which happens if all particles travel freely at the speed of light.
However, the actual particle acceleration process is expected to have a more
complicated energy dependence than assumed here. 
Therefore, the acceleration and escape time-scales derived here are only for 
instructive purposes.
We will investigate whether a more realistic energy dependence of the particle
acceleration process may change the required acceleration and escape time-scales.

%%%%%%%%%%%%%%%%%%%%%%%%%%%%%%%%%%%%%%%%%%%%%%%%%%%%%%%%%%%%%%%%%%%%%%%%%%%%%%%%
\vspace{-12pt}
\section*{Acknowledgments}

We would like to thank Fabrizio Tavecchio and Gabriele Ghisellini for providing
their BLR spectra.
This research has been supported by NASA grants 
NAG5-11796, 
NAG5-11853,
NNX12AE43G, 
\chandra GO AR9-0016X, and 
\fermi GI NNX10AO42G.
MB acknowledges support by NASA through 
\fermi GI NNX10AU11G and
ATP grant NNX10AC79G.
This work was supported in part by the Shared University Grid at Rice
funded by NSF under Grant EIA-0216467, and a partnership between Rice
University, Sun Microsystems, and Sigma Solutions, Inc.
This work was supported in part by the Data Analysis and Visualization
Cyberinfrastructure funded by NSF under grant OCI-0959097.
This work was supported in part by the Cyberinfrastructure for Computational
Research funded by NSF under Grant CNS-0821727.  
This research has made use of NASA's Astrophysics Data System and 
of the NASA/IPAC Extragalactic Database (NED) which is operated by the Jet
Propulsion Laboratory, California Institute of Technology, under contract
with the National Aeronautics and Space Administration.

%-----------------------------------------------------------------------
% \bibliography{refs_GF,refs_a-e,refs_f-k,refs_l-p,refs_q-z}

\label{lastpage}
\end{document}